\documentclass[12pt]{article}
\usepackage{amsfonts,amssymb,amsbsy,amsmath,amsthm}
\usepackage[dvips,colorlinks=true,linkcolor=blue]{hyperref}

\newtheorem{theorem}{Theorem}[section]
\newtheorem{lemma}{Lemma}[section]
\newtheorem{pr}{Proposition}[section]
\newtheorem{follow}{Corollary}[section]
\newcommand{\bel}{\begin{equation} \label}

\newcommand{\ee}{\end{equation}}

\newcommand{\rd}{{\mathbb R}^{2}}
\newcommand{\re}{{\mathbb R}}

\newcommand{\vpp}{V_{n,\omega}^{\rm per}}
\newcommand{\vp}{V^{\rm per}}
\newcommand{\vo}{V_{\omega}}
\newcommand{\npp}{{\cal N}_{n,\omega}^{\rm per}}
\newcommand{\np}{{\cal N}^{\rm per}}
\newcommand{\tr}{{\rm Tr}\,}
\newcommand{\hp}{H^{\rm per}}
\newcommand{\hpp}{H_{n,\omega}^{\rm per}}
\newcommand{\rhq}{\rho_q}
\newcommand{\rqno}{\rho_{q,n,\omega}}
\newcommand{\dq}{{\cal D}_q}
\newcommand{\cq}{{\cal C}_q}

\newcommand{\R}{\mathbb{R}}

\newcommand{\Z}{\mathbb{Z}}
\newcommand{\ze}{\mathbb{Z}}

\newcommand{\esssup}{\operatornamewithlimits{\text{ess-sup}}}

\newcommand{\pro}{\mathbb{P}}

{

\begin{document}
\begin{center}
  {\LARGE \bf Lifshitz Tails in Constant Magnetic Fields }\\
\end{center}
\begin{center}
  {\sc Fr{\'e}d{\'e}ric Klopp, Georgi Raikov}
\end{center}

\vspace{0.5cm}

{\small {\bf Abstract}: We consider the 2D Landau Hamiltonian $H$
  perturbed by a random alloy-type potential, and investigate the
  Lifshitz tails, i.e. the asymptotic behavior of the corresponding
  integrated density of states (IDS) near the edges in the spectrum of
  $H$. If a given edge coincides with a Landau level, we obtain
  different asymptotic formulae for power-like, exponential
  sub-Gaussian, and super-Gaussian decay of the one-site potential. If
  the edge is away from the Landau levels, we impose a rational-flux
  assumption on the magnetic field, consider compactly supported
  one-site potentials, and formulate a theorem which is analogous to a
  result obtained by the first author and T. Wolff in \cite{kw} in the
  case of a vanishing magnetic field.} \\

\vspace{0.5cm}

{\bf 2000 AMS Mathematics Subject Classification}: 82B44, 47B80, 47N55, 81Q10\\

\vspace{0.5cm}

{\bf Key words}: Lifshitz tails, Landau Hamiltonian, continuous Anderson model\\

\section{Introduction}
\setcounter{equation}{0} Let
\begin{equation} 
\label{1}
H_0 = H_0(b): =  (-i\nabla -A)^2 - b
\end{equation}
be the unperturbed Landau Hamiltonian, essentially self-adjoint on
$C_0^{\infty}(\rd)$. Here $A = (-\frac{bx_2}{2},\frac{bx_1}{2})$ is
the magnetic potential, and $b \geq 0$ is the constant scalar magnetic
field. It is well-known that if $b>0$, then the spectrum $\sigma(H_0)$
of the operator $H_0(b)$ consists of the so-called Landau levels
$2bq$, $q \in {\mathbb Z}_+$, and each Landau level is an eigenvalue
of infinite multiplicity. If $b = 0$, then $H_0 = -\Delta$, and
$\sigma(H_0) = [0,\infty)$ is absolutely continuous. Next, we
introduce a random ${\mathbb Z}^2$-ergodic alloy-type electric
potential $$
V(x) = V_{\omega}(x): = \sum_{\gamma \in {\mathbb Z}^2}
\omega_{\gamma} u(x - \gamma), \quad x \in \re^2.  $$
Our general
assumptions concerning the potential $V_{\omega}$ are the following
ones:
\begin{itemize}
\item ${\bf H}_1$: The single-site potential $u$ satisfies the
  estimates
  \bel{kin2}
  0 \leq u(x) \leq C_0 (1 + |x|)^{-\varkappa}, \quad x \in \re^2,
  \ee
  with some $\varkappa > 2$ and $C_0 > 0$. Moreover, there exists an
  open non-empty set $\Lambda \subset \re^2$ and a constant $C_1 > 0$
  such that $u(x) \geq C_1$ for $x \in \Lambda$.
\item ${\bf H}_2$: The coupling constants $\left\{\omega_{\gamma}\right\}_{{\gamma}
    \in {\mathbb Z}^2}$ are non-trivial, almost surely bounded i. i.
  d.  random variables.
\end{itemize}
Evidently, these two assumptions entail 
\begin{equation}
\label{0}
M : = \esssup_{\omega}\;\sup_{x \in \re^2} |\vo(x)| < \infty.  
\end{equation}
On the domain of $H_0$ define the operator $H = H_{\omega}: = H_0(b) + \vo$. The
integrated density of states (IDS) for the operator $H$ is defined as
a non-decreasing left-continuous function ${\cal N}_b : \re \to
[0,\infty)$ which almost surely satisfies
\begin{equation}
\label{2}
\int_{\re} \varphi(E) d{\cal N}_b(E) = \lim_{R \to \infty} R^{-2} 
  \tr\left({\bf 1}_{{\Lambda}_R} \varphi(H) {\bf 1}_{{\Lambda}_R} \right),
  \quad \forall \varphi \in C_0^{\infty}(\re). 
\end{equation}
Here and in the sequel ${\bf 1}_{{\cal O}}$ denotes the the
characteristic function of the set ${\cal O}$, and $\Lambda_R : =
\left(-\frac{R}{2}, \frac{R}{2}\right)^2$. By the Pastur-Shubin
formula (see e.g. \cite[Section 2]{ves} or \cite[Corollary 3.3]{hlmw}) we have
\begin{equation}
\label{10}
\int_{\re} \varphi(E) d{\cal N}_b(E) = 
{\mathbb E}\left(\tr \left({\bf 1}_{{\Lambda}_1} \varphi(H) {\bf
  1}_{{\Lambda}_1} \right) \right), \quad \forall \varphi \in C_0^{\infty}(\re), 
\end{equation}
where ${\mathbb E}$ denotes the mathematical expectation.  Moreover, there
exists a set $\Sigma \subset \re$ such that 
$\sigma(H_{\omega}) = \Sigma$ almost surely, 
and ${\rm supp}\;{d{\cal N}_b} = \Sigma$. The aim of the present article is to study
the asymptotic behavior of ${\cal N}_b$ near the edges of $\Sigma$. It is well
known that, for many random models, this behavior is characterized by a very
fast decay which goes under the name of ``Lifshitz tails''. It was studied
extensively in the absence of magnetic field (see e.g. \cite{pf}, \cite{k}),
and also in the presence of magnetic field for other types of disorder (see 
\cite{bhkl}, \cite{e1}, \cite{hlw1}, \cite{e2}, \cite{hlw2}). 
\section{Main results}
\setcounter{equation}{0}
In order to fix the picture of the almost sure spectrum  $\sigma(H_{\omega})$,
we assume $b>0$, and make the following two additional hypotheses:\\
\begin{itemize}
\item ${\bf H}_3$: The support of the random variables $\omega_{\gamma}$,
  $\gamma \in \ze^2$, consists of the interval $[\omega_-,\omega_+]$ with
  $\omega_- <\omega_+$ and $\omega_- \omega_+ \leq 0$. 

\item ${\bf H}_4$: We have $M_+ - M_- < 2b$ where 
$\pm M_{\pm} : = \esssup_{\omega}\;\sup_{x \in \re^2} \; (\pm\vo(x))$.
\end{itemize}
Assumptions ${\bf H}_1$ -- ${\bf H}_4$ imply $M_- M_+ \leq 0$.
Moreover, the union $\cup_{q=0}^{\infty} [2bq+M_-, 2bq+M_+]$ which
contains $\Sigma$, is disjoint. Introduce the bounded $\Z^2$-periodic
potential $$
W(x) : = \sum_{\gamma \in \ze^2} u(x-\gamma), \quad x \in
\re^2, $$
and on the domain of $H_0$ define the operators $H^{\pm} : =
H_0 + \omega_{\pm} W$. It is easy to see that $$
\sigma(H^-) \subseteq
\cup_{q=0}^{\infty} [2bq+M_-, 2bq], \quad \sigma(H^+) \subseteq
\cup_{q=0}^{\infty} [2bq, 2bq+M_+], $$
and $$
\sigma(H^-) \cap
[2bq+M_-, 2bq] \neq \emptyset, \quad \sigma(H^+) \cap [2bq, 2bq+M_+]
\neq \emptyset, \quad \forall q \in \Z_+.  $$
Set $$
E_q^- : =
\inf{\left\{\sigma(H^-) \cap [2bq+M_-, 2bq]\right\}}, \quad E_q^+ : =
\sup{\left\{\sigma(H^+) \cap [2bq, 2bq+M_+]\right\}}.  $$
Following
the argument in \cite{km0} (see also \cite[Theorem 5.35]{pf}), we
easily find that $$
\Sigma = \cup_{q=0}^{\infty} [E_q^-,E_q^+], $$
i.e. $\Sigma$ is represented as a disjoint union of compact intervals,
and each interval $[E_q^-,E_q^+]$ contains exactly one Landau level
$2bq$, $q \in
\Z_+$. \\
In the following theorems we describe the behavior of the integrated
density of states ${\cal N}_b$ near $E_q^-$, $q \in \ze_+$; its
behavior near $E_q^+$ could be analyzed in a completely analogous
manner. \\
Our first theorem concerns the case where $E_q^- = 2bq$, $q \in
\ze_+$. This is the case if and only if $\omega_- = 0$; in this case,
the random variables $\omega_{\gamma}$, $\gamma \in \ze^2$, are
non-negative.
\begin{theorem} \label{thin1} Let $b>0$ and assumptions ${\bf H}_1$ --
  ${\bf H}_4$ hold. Suppose that $\omega_- = 0$, and that
  \bel{fin70}
  {\mathbb P}(\omega_0 \leq E) \sim C E^{\kappa}, \quad E \downarrow
  0, \ee
  for some $C > 0$ and $\kappa > 0$. Fix the Landau level $2bq = E_q^-$, $q \in {\mathbb Z}_+$.\\
  i) Assume that $C_-(1+|x|)^{-\varkappa} \leq u(x) \leq
  C_+(1+|x|)^{-\varkappa}$, $x \in \rd$, for some $\varkappa > 2$, and
  $C_+ \geq C_- >0$. Then we have
  \begin{equation} \label{gin24} \lim_{E \downarrow 0}
    \frac{\ln{|\ln{({\cal N}_b(2bq + E) - {\cal N}_b(2bq))}|}}{\ln{E}}
    = - \frac{2}{\varkappa-2}.  \ee ii) Assume $\frac{e^{-C_+
        |x|^{\beta}}}{C_+} \leq u(x) \leq \frac{e^{-C_-
        |x|^{\beta}}}{C_-}$, $x \in \rd$, $\beta \in (0,2]$, $C_+ \geq
    C_- > 0$. Then we have
    \begin{equation} \label{gin25}
      \lim_{E \downarrow 0} \frac{\ln{|\ln{({\cal N}_b(2bq + E) - {\cal
              N}_b(2bq))|}}}{\ln{|\ln{E}|}} = 1 + \frac{2}{\beta}. 
    \end{equation}
    iii) Assume $\frac{{\bf 1}_{\{x \in \rd\,;\, |x-x_0| < \varepsilon\}}}{C_+} \leq u(x) \leq \frac{e^{-C_- |x|^2}}{C_-}$
    for some $C_+ \geq C_- > 0$, $x_0 \in \rd$, and $\varepsilon > 0$. Then there exists 
    $\delta > 0$  such that 
    $$
    1 + \delta \leq \liminf_{E \downarrow 0} \frac{\ln{|\ln{({\cal N}_b(2bq + E) -
          {\cal N}_b(2bq})|}}{\ln{|\ln{E}|}} \leq 
    $$
    \begin{equation} \label{hin2} 
      \limsup_{E \downarrow 0} \frac{\ln{|\ln{({\cal N}_b(2bq + E) -
            {\cal N}_b(2bq})|}}{\ln{|\ln{E}|}} \leq 2.
    \end{equation}
  \end{theorem}
  The proof of Theorem \ref{thin1} is contained in Sections 3 -- 5.
  In Section 3 we construct a periodic approximation of the IDS ${\cal
    N}_b$ which plays a crucial role in this proof.  The upper bounds
  of the IDS needed for the proof of Theorem \ref{thin1} are obtained
  in Section 4, and the corresponding lower bounds are deduced in
  Section 5.\\
  {\em Remarks}: i) In the first and second part of Theorem
  \ref{thin1} we consider one-site potentials $u$ respectively of
  power-like or exponential sub-Gaussian decay at infinity, and obtain
  the values of the so called Lifshitz exponents. Note however that in
  the case of power-like decay of $u$ the double logarithm of ${\cal
    N}_b(2bq + E) - {\cal N}(2bq)$ is asymptotically proportional to
  $\ln{E}$ (see \eqref{gin24}), while in the case of exponentially
  decaying $u$ this double logarithm is asymptotically proportional to
  $\ln|\ln{E}|$ (see \eqref{gin25}); in both cases the Lifshitz
  exponent is defined as the corresponding proportionality factor. In
  the third part of the theorem which deals with one-site potentials
  $u$ of super-Gaussian decay, we obtain only upper and lower bounds
  of the Lifshitz exponent. It is natural to conjecture that the value
  of this exponent is 2, i.e. that the upper bound
  in \eqref{hin2} reveals the correct asymptotic behavior.\\
  ii) In the case of a vanishing magnetic field, the Lifshitz
  asymptotics for random Schr{\"o}\-din\-ger operator with repulsive
  random alloy-type potentials has been known since long ago (see
  \cite{km}). To the authors' best knowledge the Lifshitz asymptotics
  for the Landau Hamiltonian with non-zero magnetic field, perturbed
  by a positive random alloy-type potential, is considered for the
  first time in the present article. However, it is appropriate to
  mention here the related results concerning the Landau Hamiltonian
  with repulsive random Poisson potential. In \cite{bhkl} the Lifshitz
  asymptotics in the case of a power-like decay of the one-site
  potential $u$, was investigated. The case of a compact support of
  $u$ was considered in \cite{e1}. The results for the case of a
  compact support of $u$ were essentially used in \cite{hlw1} and
  \cite{e2} (see also \cite{hlw2}),
  in order to study the problem in the case of an exponential decay of $u$.\\
  Our second theorem concerns the case where $E_q^- < 2bq$, $q \in
  \ze_+$. This is the case if and only if $\omega_- < 0$. In order to
  handle this case, we need some facts from the magnetic Floquet-Bloch
  theory.  Let $\Gamma : = g_1 \Z \oplus g_2 \Z$ with $g_j > 0$,
  $j=1,2$.  Introduce the tori \bel{kin5} {\mathbb T}_{\Gamma} : =
  \re^2/\Gamma, \quad {\mathbb T}_{\Gamma}^* : = \re^2/\Gamma^*, \ee
  where $\Gamma^* : = 2\pi g_1^{-1} \Z \oplus 2\pi g_2^{-1} \Z$ is the
  lattice dual to $\Gamma$. Denote by ${\cal O}_{\Gamma}$ and ${\cal
    O}^*_{\Gamma}$ the fundamental domains of ${\mathbb T}_{\Gamma}$
  and ${\mathbb T}_{\Gamma}^*$ respectively. Let ${\cal W} : \rd \to
  \re$ be a $\Gamma$-periodic bounded real-valued function.  On the
  domain of $H_0$
  define the operator $H_{\cal W} : = H_0 + {\cal W}$.\\
  Assume that the scalar magnetic field $b \geq 0$ satisfies {\em the
    integer-flux} condition with respect to the lattice $\Gamma$, i.e.
  that $b g_1 g_2 \in 2\pi \Z_+$.  Fix $\theta \in {\mathbb
    T}_{\Gamma}^*$. Denote by $h_0(\theta)$ the self-adjoint operator
  generated in $L^2({\cal O}_{\Gamma})$ by the closure of the
  non-negative quadratic form $$
  \int_{{\cal O}_{\Gamma}} |(i\nabla +
  A - \theta)f|^2 dx $$
  defined originally on the set $$
  \left\{f =
    g_{\big|{\cal O}_{\Gamma}} \;| \; g \in C^{\infty}(\re^2), \,
    (\tau_{\gamma}g)(x) = g(x), \, x \in \re^2, \, \gamma \in \Gamma
  \right\} $$
  where $\tau_y$, $y \in \re^2$, is the magnetic
  translation given by \bel{fin34} (\tau_y g)(x) : = e^{ib \frac{y_1
      y_2}{2}} e^{ib \frac{x\wedge y}{2}} g(x+y), \quad x \in \re^2,
  \ee with $x\wedge y : = x_1 y_2 - x_2 y_1$. Note that the
  integer-flux condition implies that the operators $\tau_{\gamma}$,
  $\gamma \in \Gamma$, commute with each other, as well as with
  operators $i\frac{\partial}{\partial x_j} + A_j$, $j=1,2$ (see
  \eqref{1}), and hence with $H_0$ and $H_{\cal W}$.  In the case
  $b=0$, the domain of the operator $h_0$ is isomorphic to the Sobolev
  space ${\rm H}^2({\mathbb T}_{\Gamma})$, but if $b>0$, this is not
  the case even under the integer-flux assumption since $h_0$ acts on
  $U(1)$-sections rather than on functions over ${\mathbb T}_{\Gamma}$
  (see e.g \cite[Subsection 2.2]{mr}).  On the domain of $h_0$ define
  the operator \bel{kin15} h_{\cal W}(\theta) : = h_0(\theta) + {\cal
    W}, \quad \theta \in {\mathbb T}_{\Gamma}^*.  \ee Set \bel{kin16}
  {\cal H}_0 : = \int_{{\cal O}_{\Gamma}^*} \oplus \; h_0(\theta)
  d\theta, \quad {\cal H}_{\cal W} : = \int_{{\cal O}_{\Gamma}^*}
  \oplus \; h_{\cal W}(\theta) d\theta.  \ee It is well-known (see e.g
  \cite{hesj}, \cite{sj}, or \cite[Subsection 2.4]{mr}) that the
  operators $H_0$ and $H_{\cal W}$ are unitarily equivalent to the
  operators ${\cal H}_0$ and ${\cal H}_{\cal W}$ respectively. More
  precisely, we have $H_0 = U^* {\cal H}_0 U$ and $H_{\cal W} = U^*
  {\cal H}_{\cal W} U$ where $U : L^2(\re^2) \to L^2({\cal O}_{\Gamma}
  \times {\cal O}_{\Gamma}^*)$ is the unitary Gelfand-type operator defined
  by
  \begin{equation} \label{uo} (Uf)(x;\theta) : = \frac{1}{\sqrt{{\rm
          vol}\,{\mathbb T}_{\Gamma}^*}} \sum_{\gamma \in \Gamma}
    e^{-i\theta(x + \gamma)}(\tau_{\gamma}f)(x), \quad x \in {\cal
      O}_{\Gamma}, \quad \theta \in {\mathbb T}_{\Gamma}^*.
  \end{equation}
  Evidently for each $\theta \in {\mathbb T}_{\Gamma}^*$ the spectrum
  of the operator $h_{\cal W}(\theta)$ is purely discrete. Denote by
  $\left\{E_j(\theta)\right\}_{j=1}^{\infty}$ the non-decreasing
  sequence of its eigenvalues.  Let $E \in \re$. Set $$
  J(E) : =
  \left\{j \in {\mathbb N} \,; \, \text{there exists} \; \theta \in
    {\mathbb T}_{\Gamma}^* \; \text{such that} \; E_j(\theta) =
    E\right\}.  $$
  Evidently, for each $E \in \re$ the set $J(E)$ is
  finite.  If $E \in \re$ is an end of an open gap in $\sigma(H_0 +
  {\cal W})$, then we will call it an edge in $\sigma(H_0 + {\cal
    W})$.  We will call the edge $E$ in $\sigma(H_0 + {\cal W})$ {\em
    simple} if $\# J(E) = 1$. Moreover, we will call the edge $E$ {\em
    non-degenerate} if for each $j \in J(E)$ the number of points
  $\theta \in {\mathbb T}_{\Gamma}^*$ such that $E_j(\theta) = E$ is
  finite, and at
  each of these points the extremum of $E_j$ is non-degenerate. \\
  Assume at first that $b = 0$. Then $H_0 = -\Delta$, and we will
  consider the general $d$-dimensional situation; the simple and
  non-degenerate edges in $\sigma(-\Delta + {\cal W})$ are defined
  exactly as in the two-dimensional case. If ${\cal W}: \re^d \to \re$
  is a real-valued bounded periodic function, it is well-known that:
  \begin{itemize}
  \item The spectrum of $-\Delta + {\cal W}$ is absolutely continuous
    (see e.g.  \cite[Theorems XIII.90, XIII.100]{rs}). In particular,
    no Floquet eigenvalue $E_j : {\mathbb T}_{\Gamma}^* \to \re$, $j
    \in {\mathbb N}$, is constant.
  \item If $d=1$, all the edges in $\sigma(-\Delta + {\cal W})$ are
    simple and non-degenerate (see e.g. \cite[Theorem XIII.89]{rs}).
  \item For $d \geq 1$ the bottom of the spectrum of $-\Delta + {\cal
      W}$ is a simple and non-degenerate edge (see \cite{ks2}).
  \item For $d \geq 1$, the edges of $\sigma(-\Delta + {\cal W})$
    generically are simple (see \cite{kr}).
  \end{itemize}
  Despite the widely spread belief that generically the higher edges
  in $\sigma(-\Delta + {\cal W})$ should also be non-degenerate in the
  multi-dimensional case $d>1$, there are no rigorous results in
  support of this
  conjecture. \\
  Let us go back to the investigation of the Lifshitz tails for the
  operator $-\Delta + V_{\omega}$. It follows from the general results
  of \cite{km0} that $E^-$ (respectively, $E^+$) is an upper
  (respectively, lower) end of an open gap in $\sigma(-\Delta +
  V_{\omega})$ if and only if it is an upper (respectively, lower) end
  of an open gap in the spectrum of $-\Delta + {\omega}_- W$
  (respectively, $-\Delta + {\omega}_+ W$). For definiteness, let us
  consider the case of an upper end $E^-$. The asymptotic behavior of
  the IDS ${\cal N}_0(E)$ as $E \downarrow E^-$ has been investigated
  in \cite{m1} - \cite{m2} in the case $d=1$, and in \cite{ks2} in the
  case $d \geq 1$ and $E^- = \inf \sigma(-\Delta + \omega_- W)$. Note
  that the proofs of the results of \cite{m1}, \cite{m2}, and
  \cite{ks2}, essentially rely on the non-degeneracy of $E^-$. Later,
  the Lifshitz tails for the operator $-\Delta + V_{\omega}$ near the
  edge $E^-$ were investigated in \cite{k} under the assumptions that
  $d \geq 1$, $E^- > \inf \sigma(-\Delta + \omega_- W)$, and that
  $E^-$ is non-degenerate edge in the spectrum of $-\Delta + \omega_-
  W$; due to the last assumption these results are conditional.
  However, it turned out possible to lift the non-degeneracy
  assumption in the two-dimensional case considered in \cite{kw}.
  First, it was shown in \cite[Theorem 0.1]{kw} that for any
  single-site potential $u$ satisfying assumption ${\bf H}_1$, we have
  $$
  \limsup_{E \downarrow 0} \frac{\ln{|\ln{({\cal N}_0(E^-+E) -
        {\cal N}_0(E_q^-))|}}}{\ln{E}} < 0 $$
  without any additional
  assumption on $E^-$. If, moreover, the support of $u$ is compact,
  and the probability $\pro(\omega_0 - \omega_- \leq E)$ admits a
  power-like decay as $E \downarrow 0$, it follows from \cite[Theorem
  0.2]{kw} that there exists $\alpha > 0$ such that \bel{kin6} \lim_{E
    \downarrow 0} \frac{\ln{|\ln{({\cal N}_0(E^-+E) - {\cal
          N}_0(E_q^-))|}}}{\ln{E}} = -\alpha \ee under the unique
  generic hypothesis that $E^-$ is a simple edge. Note that the
  absolute continuity of $\sigma(-\Delta + \omega_- W)$ plays a
  crucial role in
  the proofs of the results of \cite{kw}. \\
  Assume now that the scalar magnetic field $b>0$ satisfies the
  rational flux condition $b \in 2\pi {\mathbb Q}$. More precisely, we
  assume that $b/2\pi$ is equal to the irreducible fraction $p/r$, $p
  \in {\mathbb N}$, $r \in {\mathbb N}$. Then $b$ satisfies the
  integer-flux assumption with respect, say, to the lattice $\Gamma =
  r \Z \oplus \Z$, and the operator $H^{-}$ is unitarily equivalent to
  ${\cal H}_{\omega_- W}$. As in the non-magnetic case, in order to
  investigate the Lifshitz asymptotics as $E \downarrow E_q^-$ of
  ${\cal N}_b(E)$, we need some information about the character of
  $E_q^-$ as an edge in the spectrum of $H^-$. For example, if we
  assume that $E_q^-$ is a simple edge, and the corresponding Floquet
  band does not shrink into a point, we can repeat almost word by word
  the argument of the proof of \cite[Theorem 0.2]{kw}, and obtain the
  following
  \begin{theorem} \label{thin2} Let $b>0$, $b \in 2\pi {\mathbb Q}$,
    and assumptions ${\bf H}_1$ -- ${\bf H}_4$ hold. Assume that the
    support of $u$ is compact, $\omega_- < 0$, and $\pro(\omega_0 -
    \omega_- \leq E) \sim C E^{\kappa}$, $E \downarrow 0$, for some
    $C>0$ and $\kappa > 0$. Fix $q \in \ze_+$. Suppose $E_q^-$ is a
    simple edge in the spectrum of the operator $H^-$, and that the
    function $E_j$, $j \in J(E_q^-)$, is not identically constant.
    Then there exists $\alpha > 0$ such that \bel{hin3} \lim_{E
      \downarrow 0} \frac{\ln{|\ln{({\cal N}_b(E_q^-+E) - {\cal
            N}_b(E_q^-))|}}}{\ln{E}} = -\alpha.  \ee
  \end{theorem} {\em Remarks}: i) It is believed that under the
  rational-flux assumption the Floquet eigenvalues $E_j$, $j \in
  {\mathbb N}$, for the operator $H^-$ generically are not constant.
  Note that this property may hold only generically due to the obvious
  counterexample where $u = {\bf 1}_{\Lambda_1}$, $H^- = H_0 +
  \omega_-$, and for all $j \in {\mathbb N}$ the Floquet eigenvalue
  $E_j$ is identically equal to $2b(j-1) + \omega_-$. Also, in
  contrast to the non-magnetic case, we do not know whether the edges
  in the spectrum of $H^-$
  generically are simple.\\
  ii) The definition of the constant $\alpha$ in \eqref{hin3} is
  completely analogous to the one in \eqref{kin6} which concerns the
  non-magnetic case.  This definition involving the concepts of Newton
  polygon, Newton diagram, and Newton decay exponent, is not trivial,
  and can be found in the original work \cite{kw}, or in
  \cite[Subsection 4.2.8]{k3}.
 
\section{Periodic approximation}
\setcounter{equation}{0} Pick $a>0$ such that $\frac{ba^2}{2\pi} \in
{\mathbb N}$. Set $L : = (2n+1)/2$, $n \in {\mathbb N}$, and define
the random $2L{\mathbb Z}^2$-periodic potential $$
\vp(x) = \vpp(x): =
\sum_{\gamma \in 2L{\mathbb Z}^2}\left(\vo {\bf
    1}_{\Lambda_{2L}}\right)(x+\gamma), \quad x \in \re^2.  $$
On the
domain of $H_0$ define the operator $\hp = \hpp : = H_0 + \vpp$.  For
brevity set ${\mathbb T}_{2L} : = {\mathbb T}_{2L\Z^2}$, ${\mathbb
  T}_{2L}^* : = {\mathbb T}_{2L\Z^2}^*$ (see \eqref{kin5}).  Note that
the square $\Lambda_{2L}$ is the fundamental domain of the torus
${\mathbb T}_{2L}$, while $\Lambda_{2L}^* : = \Lambda_{\pi L^{-1}}$ is
the fundamental domain of ${\mathbb T}_{2L}^*$.  As in \eqref{kin15},
on the domain of $h_0$ define the operator $$
h(\theta) = h^{\rm
  per}(\theta): = h_0(\theta) + \vp, \quad \theta \in {\mathbb
  T}_{2L}^*, $$
and by analogy with \eqref{kin16} set $$
{\cal H}^{\rm
  per} : = \int_{\Lambda_{2L}^*} \oplus \; h^{\rm per}(\theta)
d\theta.  $$
As above, the operators $H_0$ and $\hp$ are unitarily
equivalent to the operators ${\cal H}_0$ and ${\cal H}^{\rm per}$
respectively.  Set
\begin{equation}
  \label{30}
  \np(E) = \npp(E): = (2\pi)^{-2} \int_{\Lambda_{2L}^*} N(E;h^{\rm per}(\theta)) d\theta,
  \quad E \in \re. 
\end{equation}
Here and in the sequel, if $T$ is a self-adjoint operator with purely
discrete spectrum, then $N(E; T)$ denotes the number of the
eigenvalues of $T$ less than $E \in \re$, and counted with the
multiplicities. The function $\np$ plays the role of IDS for the
operator $\hp$ since, similarly to \eqref{2} and \eqref{10}, we have
$$
\int_{\re} \varphi(E) d\np(E) = \lim_{R \to \infty} R^{-2} \tr
\left({\bf 1}_{{\Lambda}_R} \varphi(\hp) {\bf 1}_{{\Lambda}_R} \right)
$$
almost surely, and
\begin{equation}
  \label{11}
  {\mathbb E}\left(\int_{\re} \varphi(E) d\np(E)  \right) = 
  {\mathbb E}\left(\tr \left({\bf 1}_{{\Lambda}_1} \varphi(\hp) {\bf
        1}_{{\Lambda}_1} \right) \right), 
\end{equation}
for any $\varphi \in C_0^{\infty}(\re)$ (see e.g. the proof of
\cite[Theorem 5.1]{k2} where however the case of a vanishing magnetic
field is considered).
\begin{theorem} \label{t31} Assume that hypotheses ${\bf H}_1$ and
  ${\bf H}_2$ hold.  Let $q \in {\mathbb Z}_+$, $\eta > 0$. Then there
  exist $\nu > 0$ and $E_0 > 0$ such that for $E \in (0,E_0]$ and $n
  \geq E^{-\nu}$ we have $$
  {\mathbb E}\left(\np(2bq + E/2) - \np(2bq
    - E/2)\right) - e^{-E^{-\eta}} \leq {\cal N}_b(2bq + E) - {\cal
    N}_b(2bq - E) \leq $$
  \begin{equation}
    \label{26}
    {\mathbb E}\left(\np(2bq + 2E) - \np(2bq - 2E)\right) + e^{-E^{-\eta}}.
  \end{equation}
\end{theorem}
The main technical steps of the proof of Theorem \ref{t31} which is
the central result of this section, are contained in Lemmas \ref{l31}
and \ref{l32} below.
\begin{lemma} \label{l31} Let $Q = \overline{Q} \in
  L^{\infty}(\re^2)$, $X : = H_0 + Q$, $D(X) = D(H_0)$.  Then there
  exists $\epsilon = \epsilon(b) > 0$ such that for each $\alpha,
  \beta \in {\mathbb Z}^2$, and $z \in {\mathbb C} \setminus
  \sigma(X)$ we have
  \begin{equation}
    \label{3}
    \|\chi_{\alpha} (X-z)^{-1}\chi_{\beta}\|_{{\rm HS}} \leq 2\frac{b+1}{\pi^{1/2}}
    \left(1 + \frac{1}{\eta(z)}\right)\,e^{-\epsilon \eta(z) |\alpha-\beta|}
  \end{equation}
  where $\chi_{\alpha}: = {\bf 1}_{\Lambda_1 + \alpha}$, $\alpha \in
  {\mathbb Z}^2$, $\eta(z) = \eta(z;b,Q): = \frac{{\rm
      dist}(z,\sigma(X))}{|z| + |Q|_{\infty} + 1}$, $\|\cdot\|_{\rm
    HS}$ denotes the Hilbert-Schmidt norm, and $|Q|_{\infty} : =
  \|Q\|_{L^{\infty}(\re^2)}$.
\end{lemma}
\begin{proof}
  We will apply the ideas of the proof of \cite[Proposition 4.1]{k1}.
  For $\xi \in \re^2$ set $$
  X_{\xi} : = e^{\xi\cdot x} X e^{-\xi\cdot x} =
  (i\nabla + A - i\xi)^2 + Q = X - 2i\xi\cdot(i\nabla + A) + |\xi|^2.
  $$
  Evidently,
  \begin{equation}
    \label{5}
    X_{\xi} - z = (X - z) \left(1 + (X-z)^{-1} \left(|\xi|^2 - 2i\xi\cdot(i\nabla +
        A)\right)\right). 
  \end{equation}
  Let us estimate the norm of the operator $(X-z)^{-1} \left(|\xi|^2 -
    2i\xi\cdot(i\nabla + A)\right)$ appearing at the right-hand side
  of \eqref{5}. We have $$
  \|(X-z)^{-1}|\xi|^2\| \leq |\xi|^2 {\rm
    dist}(z,\sigma(X))^{-1}, $$
  $$
  \|(X-z)^{-1} 2i\xi\cdot(i\nabla +
  A)\| \leq $$
  $$
  2\|(H_0+1)^{-1}(i\nabla + A)\cdot\xi - (X-z)^{-1}
  (Q-z-1)(H_0+1)^{-1}(i\nabla + A)\cdot\xi\| \leq $$
  $$
  2C \left(1 +
    \frac{1}{\eta(z)}\right)|\xi| $$
  with $$
  C = C(b) : =
  \|(H_0+1)^{-1}(i\nabla+A)\| = \sup_{q \in {\mathbb
      Z}_+}\frac{((2q+1)b)^{1/2}}{2bq+1}.  $$
  Choose $\epsilon \in
  \left(0,\frac{1}{8(C+1)}\right)$ and $\xi \in \re^2$ such that
  $|\xi| = \epsilon \eta(z)$. Then, by the above estimates, we have $$
  \|(X-z)^{-1} \left(|\xi|^2 - 2i\xi\cdot(i\nabla + A)\right)\| \leq
  \epsilon^2 \eta(z)^2 {\rm dist}(z,\sigma(X))^{-1} + 2C\epsilon
  \left(1 + \frac{1}{\eta(z)}\right) \eta(z) \leq $$
  \begin{equation}
    \label{6}
    \epsilon^2 \eta(z) + 2C\epsilon (1 + \eta(z)) < \epsilon^2 + 4C\epsilon < 3/4
  \end{equation}
  since the resolvent identity implies $\eta(z) < 1$. Therefore, the
  operator $X_{\xi}-z$ is invertible, and
  \begin{equation}
    \label{7}
    \chi_{\alpha}(X - z)^{-1}\chi_{\beta} = \left(e^{-\xi\cdot x}
      \chi_{\alpha}\right)
    \chi_{\alpha}(X_{\xi} -z)^{-1}\chi_{\beta}  \left(e^{\xi\cdot x}
      \chi_{\beta}\right). 
  \end{equation}
  Moreover, \eqref{5} and \eqref{6} imply $$
  \|\chi_{\alpha}(X_{\xi}
  -z)^{-1}\chi_{\beta} \|_{\rm HS} \leq 4 \|(X
  -z)^{-1}\chi_{\beta}\|_{\rm HS} \leq $$
  $$
  4 \|(H_0+1)^{-1}
  \chi_{\beta} - (X -z)^{-1}(Q-z-1)(H_0+1)^{-1} \chi_{\beta}\|_{\rm
    HS} \leq $$
  \begin{equation}
    \label{8}
    4 \|(H_0+1)^{-1} \chi_{\beta}\|_{\rm HS} (1 + \|(X -z)^{-1}(Q-z-1)\|) \leq 
    4 \|(H_0+1)^{-1} \chi_{\beta}\|_{\rm HS} \left(1 + \frac{1}{\eta(z)}\right).
  \end{equation}
  Finally, applying the diamagnetic inequality for Hilbert-Schmidt
  operators (see e.g. \cite{ahs}), we get $$
  \|(H_0+1)^{-1}
  \chi_{\beta}\|_{\rm HS} \leq
  \|(H_0+1)^{-1}(H_0+b+1)\|\|(H_0+b+1)^{-1} \chi_{\beta}\|_{\rm HS}
  \leq $$
  $$
  \|(H_0+1)^{-1}(H_0+b+1)\|\|(-\Delta+1)^{-1}
  \chi_{\beta}\|_{\rm HS} = $$
  \begin{equation}
    \label{9}
    \sup_{q \in {\mathbb Z}_+} \frac{2bq + b + 1}{2bq + 1} \; \|(-\Delta+1)^{-1}
    \chi_{\beta}\|_{\rm HS} = \frac{b+1}{2\pi^{1/2}}. 
  \end{equation}
  The combination of \eqref{7}, \eqref{8}, and \eqref{9} yields $$
  \|\chi_{\alpha}(X - z)^{-1}\chi_{\beta} \|_{\rm HS} \leq
  \frac{2(b+1)}{\pi^{1/2}} e^{-\xi(\alpha-\beta)}\left(1 +
    \frac{1}{\eta(z)}\right).  $$
  Choosing $\xi = \epsilon \eta(z)
  \frac{\alpha-\beta}{|\alpha-\beta|}$, we get \eqref{3}.
\end{proof}
\begin{lemma} \label{l32} Assume that hypotheses ${\bf H}_1$ and ${\bf
    H}_2$ hold. Then there exists a constant $C>1$ such that for any
  $\varphi \in C_0^{\infty}(\re)$, and any $n \in {\mathbb N}$, $l \in
  {\mathbb N}$, we have $$
  \left|{\mathbb E}\left(\int_{\re}
      \varphi(E) d{\cal N}_b(E) - \int_{\re} \varphi(E)
      d\np(E)\right)\right| \leq $$
  \begin{equation}
    \label{12}
    c n^{-l} e^{C l \log{l}} \sup_{x\in \re, \; 0\leq j \leq l+5}
    \left|(|x|+C)^{l+5}\frac{d^j \varphi}{dx^j}(x)\right|. 
  \end{equation}
\end{lemma}
\begin{proof}
  We will follow the general lines of the proof of \cite[Lemma
  2.1]{kp}. Due to the fact that we consider only the two-dimensional
  case, and an alloy-type potential which is almost surely bounded,
  the argument here is somewhat simpler than the one in \cite{kp}. By
  \eqref{10} and \eqref{11} we have $$
  {\mathbb E}\left(\int_{\re}
    \varphi(E) d{\cal N}_b(E) - \int_{\re} \varphi(E) d\np(E)\right) =
  {\mathbb E}\left(\tr\left({\bf 1}_{\Lambda_1}(\varphi(H) -
      \varphi(\hp)) {\bf 1}_{\Lambda_1}\right)\right).  $$
  Next, we
  introduce a representation of the operator $\varphi(H) -
  \varphi(\hp)$ by the Helffer-Sj{\"o}strand formula (see e.g.
  \cite[Chapter 8]{dsj}). Let $\tilde{\varphi}$ be an almost analytic
  extension of the function $\varphi \in C_0^{\infty}(\re)$ appearing
  in \eqref{12}. We recall that $\tilde{\varphi}$ possesses the
  following properties:
  \begin{enumerate}
  \item If ${\rm Im}\;z = 0$, then $\tilde{\varphi}(z) = \varphi(z)$.
  \item ${\rm supp}\;\tilde{\varphi} \subset \{x+iy \in {\mathbb C};
    \, |y| < 1\}$.
  \item $\tilde{\varphi} \in {\cal S}\left(\{x+iy \in {\mathbb C}; \,
      |y| < 1\}\right)$.
  \item The family of functions $x \mapsto \frac{\partial
      \tilde{\varphi}}{\partial \bar{z}}(x+iy) |y|^{-m}$, $|y| \in
    (0,1)$, is bounded in ${\cal S}(\re)$ for any $m \in {\mathbb
      Z}_+$.
  \end{enumerate}
  Such extensions exist for $\varphi \in {\cal S}(\re)$ (see
  \cite{mat}, \cite[Chapter 8]{dsj}), and there exists a constant
  $C>1$ such that for any $m \geq 0$, $\alpha \geq 0$, $\beta \geq 0$,
  we have $$
  \sup_{0 \leq |y|\leq 1}\;\sup_{x \in \re}
  \left|x^{\alpha} \frac{\partial^{\beta}}{\partial
      x^{\beta}}\left(|y|^{-m} \frac{\partial
        \tilde{\varphi}}{\partial \bar{z}}(x+iy) \right)\right| \leq
  $$
  \begin{equation}
    \label{15}
    C^{m\log{m} + \alpha \log{\alpha} + \beta + 1}\sup_{\beta' \leq m + \beta + 2,
      \; \alpha' \leq \alpha} \sup_{x \in \re} \left|x^{\alpha'}
      \frac{d^{\beta'}\varphi(x)}{dx^{\beta'}}\right|. 
  \end{equation}
  Then the Helffer-Sj{\"o}strand formula yields $$
  {\mathbb
    E}\left(\tr\left({\bf 1}_{\Lambda_1}(\varphi(H) - \varphi(\hp))
      {\bf 1}_{\Lambda_1}\right)\right) = $$
  $$
  \frac{1}{\pi}{\mathbb
    E}\left(\tr\left(\int_{\mathbb C} \frac{\partial
        \tilde{\varphi}}{\partial \bar{z}}(z) \left({\bf
          1}_{\Lambda_1}\left((H-z)^{-1} - (\hp-z)^{-1}\right){\bf
          1}_{\Lambda_1}\right)dx dy\right)\right) = $$
  \begin{equation}
    \label{14a}
    \frac{1}{\pi}{\mathbb E}\left(\tr\left(\int_{\mathbb C} \frac{\partial
          \tilde{\varphi}}{\partial \bar{z}}(z) \left({\bf
            1}_{\Lambda_1}(H-z)^{-1}(\vp-V) (\hp-z)^{-1}{\bf
            1}_{\Lambda_1}\right)dx dy\right)\right). 
  \end{equation}
  Next, we will show that ${\bf 1}_{\Lambda_1}(H-z)^{-1}(\vp-V)
  (\hp-z)^{-1}{\bf 1}_{\Lambda_1}$ is a trace-class operator for $z
  \in {\mathbb C}\setminus {\mathbb R}$, and almost surely
  \begin{equation}
    \label{13}
    \|{\bf 1}_{\Lambda_1}(H-z)^{-1}(\vp-V) (\hp-z)^{-1}{\bf
      1}_{\Lambda_1}\|_{\rm Tr} \leq \frac{M(b+1)^2}{2\pi} \left(1 + \frac{M + |z| +
        1}{|{\rm Im\;z|}}\right)^2
  \end{equation}
  where $\|.\|_{\rm Tr}$ denotes the trace-class norm. Evidently, $$
  \|{\bf 1}_{\Lambda_1}(H-z)^{-1}(\vp-V) (\hp-z)^{-1}{\bf
    1}_{\Lambda_1}\|_{\rm Tr} \leq $$
  \begin{equation}
    \label{13a}
    \|{\bf 1}_{\Lambda_1}(H_0 + 1)^{-1}\|_{\rm HS}^2 \|(\vp-V)\|
    \|(H_0+1)(H-z)^{-1}\| \|(H_0+1)(\hp-z)^{-1}\|.
  \end{equation}
  By \eqref{9} we have $\|{\bf 1}_{\Lambda_1}(H_0 + 1)^{-1}\|_{\rm
    HS}^2 \leq \frac{(b+1)^2}{4\pi}$. Moreover, almost surely
  $\|\vp-V\| \leq 2M$. Finally, it is easy to check that both norms
  $\|(H_0+1)(H-z)^{-1}\|$ and $\|(H_0+1)(\hp-z)^{-1}\|$ are almost
  surely bounded from above by $1 + \frac{M + |z| + 1}{|{\rm
      Im\;z|}}$, so that  \eqref{13} follows from \eqref{13a}. Taking into
  account estimate \eqref{13} and Properties 2, 3, and 4 of the almost
  analytic continuation $\tilde{\varphi}$, we find that \eqref{14a}
  implies $$
  {\mathbb E}\left(\tr\left({\bf 1}_{\Lambda_1}(\varphi(H)
      - \varphi(\hp)) {\bf 1}_{\Lambda_1}\right)\right) = $$
  \begin{equation}
    \label{18}
    \frac{1}{\pi}\int_{\mathbb C} \frac{\partial
      \tilde{\varphi}}{\partial \bar{z}}(z) {\mathbb
      E}\left(\tr\left({\bf
          1}_{\Lambda_1}(H-z)^{-1}(\vp-V) (\hp-z)^{-1}{\bf
          1}_{\Lambda_1}\right)\right)dx dy.
  \end{equation}
  Our next goal is to obtain a precise estimate (see \eqref{17} below)
  on the decay rate as $n \to \infty$ of $$
  {\mathbb
    E}\left(\tr\left({\bf 1}_{\Lambda_1}(H-z)^{-1}(\vp-V)
      (\hp-z)^{-1}{\bf 1}_{\Lambda_1}\right)\right) $$
  with $z \in
  {\mathbb C}\setminus {\mathbb R}$ and $|{\rm Im}\;z| <1$. Evidently,
  $$
  {\mathbb E}\left(\tr\left({\bf 1}_{\Lambda_1}(H-z)^{-1}(\vp-V)
      (\hp-z)^{-1}{\bf 1}_{\Lambda_1}\right)\right) = $$
  $$
  \sum_{\alpha \in {\mathbb Z}^2, |\alpha|_{\infty} > na} {\mathbb
    E}\left(\tr\left({\bf
        1}_{\Lambda_1}\left((H-z)^{-1}\chi_{\alpha}(\vp-V)
        (\hp-z)^{-1}\right){\bf 1}_{\Lambda_1}\right)\right) $$
  where
  $|\alpha|_{\infty} : = \max_{j=1,2} |\alpha_j|$, since $\vp = V$ on
  $\Lambda_{2L}$, and therefore $\chi_{\alpha}(\vp-V) = 0$ if
  $|\alpha|_{\infty} \leq na$. Hence, bearing in mind estimates
  \eqref{0} and \eqref{3}, we easily find that $$
  |{\mathbb
    E}\left(\tr\left({\bf 1}_{\Lambda_1}(H-z)^{-1}(\vp-V) (\hp-z)^{-1}
      {\bf 1}_{\Lambda_1}\right)\right)| \leq $$
  $$
  \sum_{\alpha \in
    {\mathbb Z}^2, |\alpha|_{\infty} > na} {\mathbb E}\left(\|\chi_0
    (H-z)^{-1}\chi_{\alpha}(\vp-V) (\hp-z)^{-1}\chi_0\|_{\rm
      Tr}\right) \leq $$
  $$
  2M \sum_{\alpha \in {\mathbb Z}^2,
    |\alpha|_{\infty} > na} {\mathbb E}\left(\|\chi_0
    (H-z)^{-1}\chi_{\alpha}\|_{\rm HS}
    \|\chi_{\alpha}(\hp-z)^{-1}\chi_0\|_{\rm HS}\right) \leq $$
  \begin{equation}
    \label{14}
    \frac{M(b+1)^2}{2\pi} \left(1 + \frac{|x|+M+2}{|y|}\right)^2 
    \sum_{\alpha \in {\mathbb Z}^2,
      |\alpha|_{\infty} > na} \exp{\left(-\frac{2\epsilon |\alpha|
          |y|}{|x|+M+2}\right)}
  \end{equation}
  for every $z = x+iy$ with $0<|y|<1$. Using the summation formula for
  a geometric series, and some elementary estimates, we conclude that
  there exists a constant $C$ depending only on $\epsilon$ such that
  \begin{equation}
    \label{15a}
    \sum_{\alpha \in {\mathbb Z}^2,
      |\alpha|_{\infty} > na}
    \exp{\left(-\frac{2\epsilon |\alpha| |y|}{|x|+M+2}\right)} \leq 
    \left(1 + C\frac{|x|+M+2}{|y|}\right) \exp{\left(-\frac{a\epsilon
          n |y|}{|x|+M+2}\right)}
  \end{equation}
  provided that $0 < |y| < 1$. Putting together \eqref{14} and
  \eqref{15a}, we find that there exists a constant $C =
  C(M,b,\epsilon,a)$ such that
  \begin{equation}
    \label{16}
    \left|{\mathbb E}\left(\tr\left({\bf
            1}_{\Lambda_1}(H-z)^{-1}(\vp-V) (\hp-z)^{-1} {\bf
            1}_{\Lambda_1}\right)\right)\right| \leq
    C\left(\frac{|x|+C}{|y|}\right)^3\exp{\left(-\frac{a\epsilon n
          |y|}{|x|+C}\right)}.
  \end{equation}
  Writing $$
  \left(\frac{|x|+C}{|y|}\right)^3\exp{\left(-\frac{a\epsilon n
        |y|}{|x|+C}\right)} = (a\epsilon n)^{-l}
  \left(\frac{|x|+C}{|y|}\right)^{3+l} \left(\frac{a\epsilon
      n|y|}{|x|+C}\right)^l \exp{\left(-\frac{a\epsilon n
        |y|}{|x|+C}\right)} $$
  with $l \in {\mathbb N}$, and bearing
  in mind the elementary inequality $t^l e^{-t} \leq (l/e)^l$, $t\geq
  0$, $l \in {\mathbb N}$, we find that \eqref{16} implies $$
  \left|{\mathbb E}\left(\tr\left({\bf 1}_{\Lambda_1}(H-z)^{-1}(\vp-V)
        (\hp-z)^{-1} {\bf 1}_{\Lambda_1}\right)\right)\right| \leq $$
  \begin{equation}
    \label{17}
    C (a\epsilon e)^{-l} n^{-l}
    \left(\frac{|x|+C}{|y|}\right)^{3+l} e^{l\log{l}}, \quad l \in {\mathbb N}.
  \end{equation}
  Combining \eqref{17} and \eqref{18}, we get $$
  |{\mathbb
    E}\left(\tr\left({\bf 1}_{\Lambda_1}(\varphi(H) - \varphi(\hp))
      {\bf 1}_{\Lambda_1}\right)\right)| \leq $$
  \begin{equation}
    \label{18a}
    \frac{C}{\pi} \int_{\re} (|x|+C)^{-2}dx\;
    (a\epsilon e)^{-l} n^{-l} e^{l\log{l}} \sup_{0<|y|<1} \sup_{x \in \re}\, 
    (|x|+C)^{l+5} |y|^{-(l+3)}\left|\frac{\partial \tilde{\varphi}}{\partial
        \bar{z}}(x+iy)\right|, \quad l \in {\mathbb N}. 
  \end{equation}
  Applying estimate \eqref{15} on almost analytic extensions, we find
  that \eqref{18a} entails \eqref{12}.
\end{proof}
Now we are in position to prove Theorem \ref{t31}. Let $\varphi_+ \in
C_0^{\infty}(\re)$ be a non-negative Gevrey-class function with Gevrey
exponent $\varrho > 1$, such that $\int_{\re} \varphi_+(t)dt = 1$,
${\rm supp}\,\varphi_+ \subset \left[-\frac{E}{2},
  \frac{E}{2}\right]$. Set $\Phi_+ : = {\bf
  1}_{\left[2bq-\frac{3E}{2}, 2bq + \frac{3E}{2}\right]} * \varphi_+.$
Then $\Phi_+$ is Gevrey-class function with Gevrey exponent $\varrho$.
Moreover, $$
{\bf 1}_{\left[2bq-E, 2bq + E\right]}(t) \leq \Phi_+(t)
\leq {\bf 1}_{\left[2bq-2E, 2bq + 2E\right]}(t), \quad t \in \re.  $$
Therefore, $$
{\cal N}_b(2bq+E) - {\cal N}_b(2bq-E) \leq {\mathbb
  E}\left(\np(2bq+2E) - \np(2bq-2E)\right)\, + $$
\begin{equation}
  \label{24}
  \left|{\mathbb E}\left( \int_{\re}\Phi_+(t) d{\cal N}_b(t) - \int_{\re}
      \Phi_+(t) d\np
      (t)\right) \right|.   
\end{equation}
Applying Lemma \ref{l32} and the standard estimates on the derivatives
of Gevrey-class functions, we get
\begin{equation}
  \label{25}
  \left|{\mathbb E}\left(\int_{\re} \Phi_+(t) d{\cal N}_b(t) - \int_{\re}
      \Phi_+(t) d\np(t)\right)\right| \leq C n^{-l}
  (l+5)^{\varrho(l+5)}, \quad l \in {\mathbb N},
\end{equation}
with $C$ independent of $n$, and $l$. Optimizing the r.h.s. of
\eqref{25} with respect to $l$, we get $$
\left|{\mathbb
    E}\left(\int_{\re} \Phi_+(t) d{\cal N}_b(t) - \int_{\re} \Phi_+(t)
    d\np(t) \right)\right| \leq
\exp{\left(-(\varrho+C)n^{1/(\varrho+C)}\right)} $$
for sufficiently
large $n$. Picking $\eta > 0$, and choosing $\nu > (\varrho + C)\eta$
and $n \geq E^{-\nu}$, we find that
\begin{equation}
  \label{27}
  \left|{\mathbb E}\left(\int_{\re} \Phi_+(t) d{\cal N}_b(t) - \int_{\re}
      \Phi_+(t) d\np
      (t)\right)\right| \leq e^{-E^{-\eta}}
\end{equation}
for sufficiently small $E>0$. Now the combination of \eqref{24} and
\eqref{27} yields the upper bound in \eqref{26}. The proof of the
first inequality in \eqref{26} is quite similar, so that we will just
outline it. Let $\varphi_- \in C_0^{\infty}(\re)$ be a non-negative
Gevrey-class function with Gevrey exponent $\varrho > 1$, such that
$\int_{\re} \varphi_+(t)dt = 1$, and ${\rm supp}\,\varphi_+ \subset
\left[-\frac{E}{4}, \frac{E}{4}\right]$. Set $\Phi_+ : = {\bf
  1}_{\left[2bq-\frac{3E}{4}, 2bq + \frac{3E}{4}\right]} * \varphi_+.$
Then $\Phi_-$ is Gevrey-class function with Gevrey exponent $\varrho$.
Similarly to \eqref{24} we have $$
{\mathbb E}\left(\np(2bq+E/2) -
  \np(2bq-E/2)\right) - \left|\int_{\re}{\mathbb E}\left( \Phi_-(t)
    d{\cal N}_b(t) - \int_{\re} \Phi_-(t) d\np (t)\right)\right| \leq
$$
\begin{equation}
  \label{28}
  \leq {\cal N}_b(2bq+E) - {\cal N}_b(2bq-E).    
\end{equation}
Arguing as in the proof of \eqref{27}, we obtain $$
\left|\int_{\re}
  {\mathbb E}\left( \Phi_-(t) d{\cal N}_b(t) - \int_{\re} \Phi_-(t)
    d\np (t)\right) \right| \leq e^{-E^{-\eta}} $$
which combined with
\eqref{28} yields the lower bound
in~\eqref{26}. Thus, the proof of Theorem \ref{t31} is now complete.\\
Further, we introduce a reduced IDS $\rhq$ related to a fixed Landau
level $2bq$, $q \in {\mathbb Z}_-$. \\
It is well-known that for every fixed $\theta \in {\mathbb T}_{2L}^*$
we have $\sigma(h(\theta)) = \cup_{q=0}^{\infty} \left\{2bq\right\}$,
and ${\rm dim \; Ker}\;(h(\theta) - 2bq) = 2bL^2/\pi$ for each $q \in
{\mathbb Z}_+$ (see \cite{duno}). Denote by $p_q(\theta) :
L^2(\Lambda_{2L}) \to L^2(\Lambda_{2L})$ the orthogonal projection
onto ${\rm Ker}\;(h(\theta) - 2bq)$, and by $r_q(\theta) =
r_{q,n,{\omega}}(\theta)$ the operator $p_q(\theta) \vpp p_q(\theta)$
defined and self-adjoint on the finite-dimensional Hilbert space
$p_q(\theta) L^2(\Lambda_{2L})$. Set
\begin{equation} \label{rids} \rhq(E) = \rqno(E) = (2\pi)^{-2}
  \int_{\Lambda_{2L}^*} N(E;r_{q,n,\omega}(\theta)) d\theta, \quad E
  \in \re.
\end{equation}
By analogy with \eqref{30}, we call the function $\rqno$ the IDS for
the operator ${\cal R}_q = {\cal R}_{q,n,\omega} : =
\int_{\Lambda_{2L}^*} \oplus r_{q,n,\omega} d\theta$ defined and
self-adjoint on ${\cal P}_q L^2(\Lambda_{2L} \times \Lambda_{2L}^*)$ where
${\cal P}_q : = \int_{\Lambda_{2L}^*} \oplus p_q(\theta) d\theta$.
Note that ${\cal R}_q = {\cal P}_q \vp {\cal P}_q$.\\
Denote by $P_q$, $q \in {\mathbb Z}_+$, the orthogonal projection onto
${\rm Ker}(H_0 - 2bq)$. Evidently, ${\cal P}_q = U P_q U^*$.  As
mentioned in the Introduction, ${\rm rank}\,P_q = \infty$ for every $q
\in {\mathbb Z}_+$.  Moreover, the functions
\begin{equation} \label{br1} e_{j}(x) = e_{j,q}(x) : = (-i)^q
  \sqrt{\frac{q!}{\pi j!}}  \left(\frac{b}{2}\right)^{(j-q+1)/2} (x_1
  + ix_2)^{j-q} L_q^{(j-q)}\left(\frac{b}{2}|x|^2\right)
  e^{-\frac{b}{4}|x|^2}, \, j \in {\mathbb Z}_+,
\end{equation}
form the so-called angular-momentum orthogonal basis of $P_q
L^2(\rd)$, $q \in {\mathbb Z}_+$ (see \cite{f} or \cite[Section
9]{bpr}). Here $$
L_q^{(j-q)}(\xi) : = \sum_{l={\rm max}\{0,q-j\}}^q
\frac{j!}{(j-q+l)!(q-l)!}  \frac{(-\xi)^l}{l!}, \quad \xi \in \re,
\quad q \in {\mathbb Z}_+, \quad j \in {\mathbb Z}_+, $$
are the
generalized Laguerre polynomials. For further references we give here
several estimates concerning the functions $e_{j,k}$. If $q \in
{\mathbb Z}_+$, $j \geq 1$, and $\xi \geq 0$, we have
\begin{equation} \label{mrs1} L_q^{(j-q)}(j\xi)^2 \leq j^{2q} e^{2\xi}
\end{equation} 
(see \cite[Eq. (4.2)]{hlw}). On the other hand, there exists $j_0 > q$
such that $j \geq j_0$ implies
\begin{equation} \label{mrs2} L_q^{(j-q)}(j\xi)^2 \geq
  \frac{1}{(q!)^2} \left(\frac{1}{2}\right)^{2+2q} (j-q)^{2q}
\end{equation}
if $\xi \in [0,1/2]$ (see \cite[Eq. (3.6)]{rw}). Moreover, for $j \in
\ze_+$ and $q \in \ze_+$ we have \bel{fin11} e_{j,q}(x) =
\frac{1}{\sqrt{q!(2b)^q}} (a^*)^q e_{0,q}(x), \quad x \in \re, \ee
where \bel{fin10} a^* : = -i \frac{\partial}{\partial x_1} - A_1
-i\left(-i \frac{\partial}{\partial x_2} - A_2\right) = -2i
e^{b|z|^2/4} \frac{\partial}{\partial z} e^{-b|z|^2/4}, \quad z : =
x_1 + i x_2, \ee is the creation operator (see e.g. \cite[Section
9]{bpr}). Evidently, $a^*$ commutes with the magnetic translation
operators $\tau_{\gamma}$, $\gamma \in 2L\ze^2$ (see \eqref{fin34}).
Finally, the projection $P_q$, $q \in {\mathbb Z}_+$, admits the
integral kernel
\begin{equation} \label{kqb} K_{q,b}(x,x') = \frac{b}{2\pi}
  e^{-i\frac{b}{2} x \wedge x'}
  \Psi_q\left(\frac{b}{2}|x-x'|^2\right), \quad x, x' \in \rd,
\end{equation}  
where $\Psi_q(\xi) : = L^{(0)}_q(\xi) e^{- \xi/2}$, $\xi \in \re$.
Since $P_q$ is an orthogonal projection in $L^2(\rd)$ we have
$\|P_q\|_{L^2(\rd) \to L^2(\rd)} = 1$.  Using the facts that ${\cal
  P}_q = U P_q U^*$ and ${\cal P}_q : = \int_{\Lambda_{2L}^*} \oplus
p_q(\theta) d\theta$, as well as the explicit expressions \eqref{uo}
for the unitary operator $U$, and \eqref{kqb} for the integral kernel
of $P_q$, $q \in {\mathbb Z}_+$, we easily find that the projection
$p_q(\theta)$, $\theta \in {\mathbb T}_{2L}^*$, admits an explicit
kernel in the form $$
{\cal K}_{q,b}(x,x';\theta) = \frac{b}{2\pi}
e^{i\theta(x'-x)} e^{-i\frac{b}{2} x \wedge x'} \times $$
\begin{equation} \label{kqbt} \sum_{\alpha \in 2L{\mathbb Z}^2}
  \Psi_q\left(\frac{b}{2}|x-x'+\alpha|^2\right) e^{-i\theta \alpha}
  e^{i{\frac{b}{2} (x+x') \wedge \alpha}} e^{i{\frac{b}{2} \alpha_1
      \alpha_2}}, \quad x, x' \in \Lambda_{2L}.
\end{equation}
\begin{lemma} \label{l33} Let the assumptions of Theorem~\ref{t31}
  hold. Suppose, moreover, that
  the random variables $\omega_\gamma$, $\gamma \in {\mathbb Z}^2$, are non-negative.\\
  a) For each $c_0 \in \left(1 + \frac{M}{2b},\infty\right)$ there
  exists $E_0 \in (0,2b)$ such that for each $E \in (0,E_0)$, $\theta
  \in {\mathbb T}_{2L}^*$, almost surely
  \begin{equation} \label{31} N(E;r_0(\theta)) \leq N(E; h(\theta))
    \leq N(c_0 E; r_0(\theta)).
  \end{equation}
  b) Assume ${\bf H}_4$, i.e. $2b > M$. Then for each $c_1 \in
  \left(0,1 - \frac{M}{2b}\right)$, $c_2 \in \left(1 +
    \frac{M}{2b},\infty\right)$, there exists $E_0 \in (0,2b)$ such
  that for each $E \in (0,E_0)$, $\theta \in {\mathbb T}_{2L}^*$, and
  $q \geq 1$, almost surely
  \begin{equation}
    \label{32}
    N(c_1 E; r_q(\theta)) \leq N(2bq + E; h(\theta)) - N(2bq;
    h(\theta)) \leq N(c_2 E; r_q(\theta)).
  \end{equation}
\end{lemma}
\begin{proof}
  In order to simplify the notations we will omit the explicit
  dependence of the operators $h$, $h_0$, $p_q$, and $r_q$, on
  $\theta\in {\mathbb T}_{2L}^*$. Moreover, we set $\dq : = p_q D(h) =
  p_q L^2(\Lambda_{2L})$, and $\cq : = (1-p_q) D(h)$.  At first we
  prove \eqref{31}. The minimax principle implies $$
  N(E;h) \geq N(E;
  {p_0 h p_0}_{|{\cal D}_0}) = N(E; r_0) $$
  which coincides with the
  lower bound in \eqref{31}. On the other hand, the operator
  inequality
  \begin{equation}
    \label{34}
    h \geq p_0 (h_0 + (1-\delta)\vp) p_0 + (1-p_0)(h_0 + (1-\delta^
    {-1})\vp)(1-p_0),
    \quad \delta \in (0,1),  
  \end{equation}
  combined with the minimax principle, entails
  \begin{equation}
    \label{33}
    \begin{split}
      N(E;h) &\leq N(E; {p_0 (h_0 + (1-\delta)\vp) p_0}_{|{\cal D}_0})
      \\ &\hskip3cm + N(E;(1-p_0)(h_0 + (1-\delta^
      {-1})\vp)(1-p_0)_{|{\cal C}_0})\\
      &\leq N((1-\delta)^{-1}E; r_0) + N(E +
      M(\delta^{-1}-1);(1-p_0)h_0(1-p_0)_{|{\cal C}_0}).
    \end{split}
  \end{equation}
  Choose $M(\delta^{-1}-1) < 2b$, and, hence, $c_0 : = (1-\delta)^{-1}
  > 1 + \frac{M}{2b}$, and $E \in (0, 2b - M(\delta^{-1}-1))$. Since
  $$
  \inf \sigma((1-p_0)h_0(1-p_0)_{|{\cal C}_0}) = 2b, $$
  we find that
  the second term on the r.h.s. of \eqref{33} vanishes, and $N(E; h)
  \leq N(c_0 E; r_0)$ which coincides with the upper bound in
  \eqref{31}. \\
  Next we assume $q \geq 1$ and $M < 2b$, and prove \eqref{32}. Note
  for any $E_1 \in (0,2b-M)$ we have $$
  N(2bq; h) = N(2bq - E_1; h).
  $$
  Pick again $\delta \in \left(\frac{M}{2b+M},0\right)$ so that
  $c_2 : = (1-\delta)^{-1} > 1 + \frac{M}{2b}$. Then the operator
  inequality $$
  h \geq p_q (h_0 + (1-\delta)\vp) p_q + (1-p_q)(h_0 +
  (1-\delta^ {-1})\vp)(1-p_q), \quad \delta \in (0,1), $$
  analogous to
  \eqref{34}, yields
  \begin{equation*}
    \begin{split}
      N(2bq + E;h) &\leq N(2bq + E; {p_q (h_0 + (1-\delta)\vp)
        p_q}_{|\dq})\\ &\hskip3cm + N(2bq + E;(1-p_q)(h_0 + (1-\delta^
      {-1})\vp)(1-p_q)_{|\cq}) \\&\leq N(c_2 E; r_q) + N(2bq + E +
      M(\delta^{-1}-1);(1-p_q)h_0(1-p_q)_{|{\cq}}).
    \end{split}
  \end{equation*}
  On the other
  hand, the minimax principle implies $$
  N(2bq - E_1; h) \geq
  N(2bq-E_1; (1-p_q) h (1-p_q)_{|\cq}) \geq N(2bq-E_1-M; (1-p_q) h_0
  (1-p_q)_{|{\cq}}).  $$
  Thus we get
  \begin{equation}
    \label{35}
    \begin{split}
      N(2bq + E; h) &- N(2bq-E_1; h) \leq N(c_2 E; r_q) \\
      &\hskip2cm + N(2bq + E +
      M(\delta^{-1}-1);(1-p_q)h_0(1-p_q)_{|\cq}) \\ &\hskip2cm -
      N(2bq-E_1-M; (1-p_q)h_0 (1-p_q)_{|\cq}).
    \end{split}
  \end{equation}
  It is easy to check that
  $$
  2bq - E_1 - M > 2b(q-1), \quad 2bq + E +
  M(\delta^{-1}-1) < 2(q+1)b $$
  provided that $E \in (0, 2b
  -M(\delta^{-1} -1))$. Since $$
  \sigma((1-p_q)h_0(1-p_q)_{|\cq}) \cap
  (2(q-1)b, 2(q+1)b) = \emptyset, $$
  we find that the the r.h.s. of
  \eqref{35} is equal to $N(c_2 E; r_q)$, thus
  getting the upper bound in \eqref{32}. \\
  Finally, we prove the lower bound in \eqref{32}. Pick $\zeta \in
  \left(\frac{M}{2b-M},\infty\right)$, and, hence $c_1 : =
  (1+\zeta)^{-1} \in \left(0,\frac{M}{2b}\right)$.  Bearing in mind
  the operator inequality $$
  h \leq p_q (h_0 + (1+\zeta)\vp) p_q +
  (1-p_q)(h_0 + (1+\zeta^ {-1})\vp)(1-p_q), $$
  and applying the
  minimax principle, we obtain 
  \begin{equation*}
    \begin{split}
  N(2bq + E;h) &\geq 
  N(2bq + E;
  {p_q (h_0 + (1+\zeta)\vp) p_q}_{|\dq})
  \\ &\hskip3cm+ N(2bq + E;(1-p_q)(h_0 +
  (1+\zeta^ {-1})\vp)(1-p_q)_{|\cq})\\ &\geq 
  N(c_1 E; r_q) + N(2bq
  + E - M(\zeta^{-1}+1);(1-p_q)h_0(1-p_q)_{|\cq}).      
    \end{split}
  \end{equation*}
  On the other
  hand, since $\vp \geq 0$, the minimax principle directly implies $$
  N(2bq-E_1;h) \leq N(2bq-E_1;h_0) = N(2bq-E_1; (1-p_q)h_0
  (1-p_q)_{|\cq}).  $$
  Combining the above estimates, we get 
  \begin{multline}
    \label{36}
  N(2bq + E; h) - N(2bq-E_1; h) \geq N(c_1 E; r_q) \\
    -\left|N(2bq + E - M(\zeta^{-1}+1);(1-p_q)h_0(1-p_q)_{|\cq})\right.
        \\ \left.-
      N(2bq-E_1; (1-p_q) h_0 (1-p_q)_{|{\cq}})\right|. 
  \end{multline}
  Since $$
  2(q-1)b < 2bq + E - M(\zeta^{-1}+1) < 2(q+1)b, \quad
  2(q-1)b < 2bq - E_1 < 2(q+1)b, $$
  provided that $E \in (0, 2b +
  M(\zeta^{-1}+1))$, we find that the r.h.s of \eqref{36} is equal to
  $N(c_1 E; r_q)$ which entails the lower bound in \eqref{32}.
\end{proof}
Integrating \eqref{31} and \eqref{32} with respect to $\theta$ and
$\omega$, and combining the results with \eqref{26}, we obtain the
following
\begin{follow} \label{f31} Assume that the hypotheses of Theorem
  \ref{t31} hold.  Let $q \in {\mathbb Z}_+$ $\eta > 0$. If $q\geq 1$,
  assume $M<2b$. Then there exist $\nu = \nu(\eta) > 0$, $d_1 \in
  (0,1)$, $d_2 \in (1,\infty)$, and $\tilde{E}_0 > 0$, such that for
  each $E \in (0,\tilde{E}_0)$ and $n \geq E^{-\nu}$, we have
  \begin{equation}
    \label{37}
    {\mathbb E}\left(\rqno(d_1 E)\right) - e^{-E^{-\eta}} \leq {\cal N}_b(2bq + E) -
    {\cal N}_b(2bq) \leq {\mathbb E}\left(\rqno(d_2 E)\right) + e^{-E^{-\eta}}.
  \end{equation}
\end{follow}
\section{Proof of Theorem \ref{thin1}: upper bounds of the IDS}
\setcounter{equation}{0} In this section we obtain the upper bounds of
${\cal N}_b(2bq+E) - {\cal N}_b(2bq)$ necessary for the proof of Theorem
\ref{thin1}.
\begin{theorem} \label{t21} Assume that ${\bf H}_1$ -- ${\bf H}_4$
  hold, that almost surely $\omega_{\gamma} \geq 0$, $\gamma \in
  {\mathbb Z}^2$, and \eqref{fin70} is valid.
  Fix the Landau level $2bq$, $q \in {\mathbb Z}_+$.\\
  i) Assume that $u(x) \geq C(1 + |x|)^{-\varkappa}$, $x \in \rd$, for
  some $\varkappa > 2$, and $C>0$. Then we have
  \begin{equation} \label{gin22} \liminf_{E \downarrow 0}
    \frac{\ln{|\ln{({\cal N}_b(2bq + E) - {\cal
            N}_b(2bq))|}}}{|\ln{E}|} \geq \frac{2}{\varkappa-2}.  \ee
    ii) Assume $u(x) \geq C e^{-C|x|^{\beta}}$, $x \in \rd$, for some
    $\beta > 0$, $C>0$. Then we have
    \begin{equation} \label{br22}
      \liminf_{E \downarrow 0} \frac{\ln{|\ln{({\cal N}_b(2bq + E) - {\cal
              N}_b(2bq))|}}}{\ln{|\ln{E}|}} \geq 1 + \frac{2}{\beta}. 
    \end{equation}
    iii) Assume $u(x) \geq C {\bf 1}_{\{x \in \rd\,;\, |x-x_0| < \varepsilon\}}$
    for some $C > 0$, $x_0 \in \rd$, and $\varepsilon > 0$. Then there exists 
    $\delta > 0$  such that we have 
    \begin{equation} \label{br23}
      \liminf_{E \downarrow 0} \frac{\ln{|\ln{({\cal N}_b(2bq + E) - {\cal N}_b(2bq})|}}{\ln{|\ln{E}|}} \geq 1 + \delta. 
    \end{equation}
  \end{theorem}
  Fix $\theta \in {\mathbb T}_{2L}^*$. Denote by $\lambda_j(\theta)$,
  $j = 1, \ldots, {\rm rank}\,r_{q,n,\omega}(\theta)$, the eigenvalues
  of the operator $r_{q,n,\omega}(\theta)$ enumerated in
  non-decreasing order. Then \eqref{rids} implies
  \begin{equation} \label{br7} {\mathbb E}\left(\rqno(E)\right) =
    \frac{1}{(2\pi)^2} \int_{\Lambda_{2L}^*} {\mathbb E}(N(E;
    r_{q,n,\omega}(\theta)) d\theta = \frac{1}{(2\pi)^2}
    \int_{\Lambda_{2L}^*} \sum_{j=1}^{{\rm
        rank}\,r_{q,n,\omega}(\theta)} {\mathbb P}(\lambda_j(\theta) <
    E) d\theta
  \end{equation} 
  with $E \in \re$.  Since the potential $V$ is almost surely bounded,
  we have ${\rm rank}\,r_{q,n,\omega}(\theta) \leq {\rm
    rank}\,p_{q}(\theta) = 2bL^2/\pi$. Therefore, \eqref{br7} entails
  \begin{equation} \label{br8} {\mathbb E}\left(\rqno(E)\right) \leq
    \frac{bL^2}{2\pi^3} \int_{\Lambda_{2L}^*}
    \pro(r_{q,n,\omega}(\theta)\; \text{has an eigenvalue less
      than}\;E) d\theta.
  \end{equation}
  In order to estimate the probability in \eqref{br8}, we need the
  following
  \begin{lemma}
    \label{le:1}
    Assume that, for $n\sim E^{-\nu}$, the operator $r_{q,
      n,\omega}(\theta)$ has an eigenvalue less than $E$. Set $L : =
    (2n+1)a/2$. Pick $E$ small and $l$ large such that $L >> l$ both
    large. Decompose $\Lambda_{2L}=\cup_{\gamma\in
      2l\Z^2\cap\Lambda_{2L}}(\gamma+\Lambda_{2l})$.  Fix ${\cal C} >
    1$ sufficiently large and $m=m(L,l)$ such that
    \begin{gather}
      \label{451} \frac{1}{{\cal C}} bl^2 \leq m \leq
      {\cal C} bL^2,
    \\ \label{452} E\left(\frac{l}{L}\right)^2 > {\cal
        C} e^{-bl^2/2+m\ln({\cal C} bl^2/m)}.
    \end{gather}
    Then, there exists $\gamma\in 2l\Z^2\cap\Lambda_{2L}$ and a non
    identically vanishing function $\psi \in L^2(\rd)$ in the span of
    $\{e_{j,q}\}_{0\leq j\leq m}$, the functions $e_{j,q}$ being
    defined in \eqref{br1}, such that
    \begin{equation}
      \label{eq:11}
      \langle V^\gamma_\omega\psi,\psi\rangle_l\leq
      2E\langle\psi,\psi\rangle_l
    \end{equation}
    where $V^\gamma_\omega(x)=V^{\rm per}_\omega(x+\gamma)$, and
    $\langle\cdot,\cdot\rangle_l : = \int_{\Lambda_{2l}}|\cdot|^2 dx$.
  \end{lemma}
  \begin{proof}
    Consider $\varphi\in {\rm Ran}\,p_q(\theta)$ a normalized
    eigenfunction of the operator $r_{q,n,\omega}(\theta)$
    corresponding to an eigenvalue smaller than $E$.  Then we have
    \begin{equation}
      \label{eq:28}
      \langle V_\omega\varphi,\varphi\rangle_L \leq E\langle\varphi,\varphi\rangle_L.
    \end{equation}
    Whenever necessary, we extend $\varphi$ by magnetic periodicity
    (i.e. the periodicity with respect to the magnetic translations)
    to the whole plane $\rd$. Note that $$
    \varphi(x) =
    \varphi(x;\theta) = \int_{\Lambda_{2L}} {\cal
      K}_{q,b}(x,x';\theta) \varphi(x') dx' = \frac{b}{2\pi}
    \int_{\rd} e^{i\theta(x'-x)} K_{q,b}(x,x') \varphi(x') dx' $$
    with
    $x \in \Lambda_{2L}$ (see \eqref{kqb} and \eqref{kqbt} for the definition
    of $K_{q,b}$ and ${\cal
      K}_{q,b}$ respectively). Evidently, $\varphi \in L^{\infty}(\rd)$,
    and since it is normalized in $L^2(\Lambda_{2L})$, we have $$
    \|\varphi\|_{L^{\infty}(\rd)} \leq \sup_{x \in \Lambda_{2L}}
    \left(\int_{\Lambda_{2L}} |{\cal K}_{q,b}(x,x';\theta)|^2
      dx'\right)^{1/2} \leq $$
    \bel{gin1} \sup_{x \in \Lambda_{2L}}
    \left(\int_{\Lambda_{2L}}\left(\sum_{\alpha \in 2L{\mathbb Z}^2}
        \tilde{\Psi}_q(x-x'+\alpha)\right)^2 dx'\right)^{1/2} \leq C
    \ee where
    \begin{equation} \label{br6} \tilde{\Psi}_q(y) : = \frac{b}{2\pi}
      \left|\Psi_q\left(\frac{b}{2}|y|^2\right)\right|, \quad y \in
      \rd.
    \end{equation}
    and $C$ depends on $q$ and $b$ but is independent of $n$ and $\theta$.\\
    Fix $C_1>1$ large to be chosen later on. Consider the sets
    \begin{gather*}
      \mathcal{L}_+=\left\{\gamma\in 2l\Z^2\cap\Lambda_{2L};\
        \int_{\gamma+\Lambda_{2l}}|\varphi(x)|^2dx\geq
        \frac1{C_1}\left(\frac{l}{L}\right)^2
        \int_{\Lambda_{2L}}|\varphi(x)|^2dx\right\},\\
      \mathcal{L}_-=\left\{\gamma\in 2l\Z^2\cap\Lambda_{2L};\
        \int_{\gamma+\Lambda_{2l}}|\varphi(x)|^2dx<
        \frac1{C_1}\left(\frac{l}{L}\right)^2
        \int_{\Lambda_{2L}}|\varphi(x)|^2dx\right\}.
    \end{gather*}
    The sets $\mathcal{L}_-$ and $\mathcal{L}_+$ partition
    $2l\Z^2\cap\Lambda_{2L}$.\\
    Fix $C_2>1$ large. Let us now prove that for some $\gamma\in
    \mathcal{L}_+$, one has
    \begin{equation}
      \label{eq:2}
      \int_{\gamma+\Lambda_{2l}}V^{\rm per}_\omega(x)|\varphi(x)|^2dx\leq C_2 E
      \int_{\gamma+\Lambda_{2l}}|\varphi(x)|^2dx. 
    \end{equation}
    Indeed, if this were not the case, then~\eqref{eq:28} would yield
    \begin{equation}
      \label{eq:3}
      \begin{split}
        -E\sum_{\gamma\in\mathcal{L}_-}\int_{\gamma+\Lambda_{2l}}|\varphi(x)|^2dx
        &\leq\sum_{\gamma\in\mathcal{L}_-}
        \left(\int_{\gamma+\Lambda_{2l}}V^{\rm
            per}_\omega(x)|\varphi(x)|^2dx-
          E\int_{\gamma+\Lambda_{2l}}|\varphi(x)|^2dx\right)\\
        &\leq \sum_{\gamma\in\mathcal{L}_+}
        \left(E\int_{\gamma+\Lambda_{2l}}|\varphi(x)|^2dx-
          \int_{\gamma+\Lambda_{2l}}V^{\rm
            per}_\omega(x)|\varphi(x)|^2dx
        \right)\\
        &\leq -E(C_2-1)\sum_{\gamma\in\mathcal{L}_+}
        \int_{\gamma+\Lambda_{2l}}|\varphi(x)|^2dx. 
      \end{split}
    \end{equation}
    On the other hand, the definition of $\mathcal{L}_-$ yields
    \begin{equation*}
      \begin{split}
        \int_{\Lambda_{2L}}|\varphi(x)|^2dx&=
        \sum_{\gamma\in\mathcal{L}_-}\int_{\gamma+\Lambda_{2l}}|\varphi(x)|^2dx
        +\sum_{\gamma\in\mathcal{L}_+}\int_{\gamma+\Lambda_{2l}}|\varphi(x)|^2dx\\
        &\leq
        \sum_{\gamma\in\mathcal{L}_+}\int_{\gamma+\Lambda_{2l}}|\varphi(x)|^2dx
        +\frac{1}{C_1}\sum_{\gamma\in\mathcal{L}_-}
        \left(\frac{l}{L}\right)^2\int_{\Lambda_{2L}}|\varphi(x)|^2dx\\
        &\leq
        \sum_{\gamma\in\mathcal{L}_+}\int_{\gamma+\Lambda_{2l}}|\varphi(x)|^2dx
        +\frac{1}{C_1}\int_{\Lambda_{2L}}|\varphi(x)|^2dx. 
      \end{split}
    \end{equation*}
    Plugging this into~\eqref{eq:3}, we get
    \begin{equation}
      \label{eq:4}
      \frac{E}{C_1}  \int_{\Lambda_{2L}}|\varphi(x)|^2dx\geq E(C_2-1)
      \left(1-\frac{1}{C_1}\right)\int_{\Lambda_{2L}}|\varphi(x)|^2dx
    \end{equation}
    which is clearly impossible if we choose $(C_2-1)(C_1-1)>1$.\\
    So from now on we assume that $(C_2-1)(C_1-1)>1$. Hence, we can
    find $\gamma\in 2l\Z^2\cap\Lambda_{2L}$ such that one has
$$
\int_{\gamma+\Lambda_{2l}}V^{\rm per}_\omega(x)|\varphi(x)|^2dx\leq
C_2 E \int_{\gamma+\Lambda_{2l}}|\varphi(x)|^2dx, $$
$$
\int_{\gamma+\Lambda_{2l}}|\varphi(x)|^2dx\geq
\frac1{C_1}\left(\frac{l}{L}\right)^2
\int_{\Lambda_{2L}}|\varphi(x)|^2dx.  $$
Shifting the variables in the integrals above by $\gamma$, we may
assume $\gamma=0$ if we replace $V^{\rm per}_\omega$ by
$V^\gamma_\omega$. Thus  we get
$$
  \int_{\Lambda_{2l}}V^{\gamma}_\omega(x)|\varphi(x)|^2dx\leq C_2 E
  \int_{\Lambda_{2l}}|\varphi(x)|^2dx,
$$
$$
  \int_{\Lambda_{2l}}|\varphi(x)|^2dx\geq
  \frac1{C_1}\left(\frac{l}{L}\right)^2
  \int_{\gamma+\Lambda_{2L}}|\varphi(x)|^2dx.
$$
Due to the magnetic periodicity of $\varphi$, we have 
\begin{equation*}
  \int_{\gamma+\Lambda_{2L}}|\varphi(x)|^2dx=\int_{\Lambda_{2L}}|\varphi(x)|^2dx
\end{equation*}
which yields
\begin{equation}
  \label{411a}
  \int_{\Lambda_{2l}}V_\omega(x)|\varphi(x)|^2dx\leq C_2 E
  \int_{\Lambda_{2l}}|\varphi(x)|^2dx, 
\end{equation}
\begin{equation} \label{411b}
  \int_{\Lambda_{2l}}|\varphi(x)|^2dx\geq
  \frac1{C_1}\left(\frac{l}{L}\right)^2
  \int_{\Lambda_{2L}}|\varphi(x)|^2dx.
\end{equation}
Let us now show that roughly the same estimates hold true for
$\varphi$ replaced by a function $\psi \in P_q L^2(\R^2)$.  Set $\psi
: = P_q \chi_- e_{\theta} \varphi$ where $e_{\theta}(x) : = e
^{i\theta x}$, $x \in \rd$, and $\chi_-$ denotes the characteristic
function of the set $\{x \in \rd; |x|_{\infty} < L\}$.  Note that
$\varphi - \overline{e_{\theta}} \psi = \overline{e_{\theta}} P_q
\chi_+ e_{\theta} \varphi$ where $\chi_+$ is the characteristic
function of the set $\{x \in \rd; |x|_{\infty} \geq L\}$. Let us
estimate the $L^2(\Lambda_{2L})$-norm of the function $\varphi -
\overline{e_{\theta}} \psi$. We have 
\begin{equation}
  \label{br28}
  \begin{split}
    \|\varphi - \overline{e_{\theta}}\psi\|^2_L &: = \|\varphi -
    \overline{e_{\theta}}\psi\|^2_{L^2(\Lambda_{2L})} \\ &=
    \int_{\Lambda_{2L}} \left|\int_{\rd}
      K_{q,b}(x,x')\chi_+(x')e^{i\theta x} \varphi(x') dx'\right|^2 dx
    \\ &\leq\sup_{x' \in \rd} |\varphi(x')|^2 \int_{\Lambda_{2L}}
    \int_{\rd} \int_{\rd} \tilde{\Psi}_q(x-x') \tilde{\Psi}_q(x-x'')
    \chi_+(x') \chi_+(x'') dx' dx'' dx,
  \end{split}
\end{equation}
the function $\tilde{\Psi}$ being defined in \eqref{br6}.
Bearing in mind estimate \eqref{gin1}, and taking into account the Gaussian
decay of $\tilde{\Psi}$ at infinity, we easily find that \eqref{br28} implies the existence of a constant $C>0$ such that for sufficiently large $L$ we have 
\begin{equation*}
\|\varphi - \overline{e_{\theta}}\psi\|^2_L \leq e^{-L/C}.
\end{equation*}
 As $\varphi$
is normalized in $L^2(\Lambda_{2L})$, this implies that, for
sufficiently small $E$,
\begin{equation} \label{br26}
  \|\psi\|_L\geq\frac12\|\varphi\|_L
  \quad\text{and}\quad
  \|\varphi - \overline{e_{\theta}}\psi\|_L\leq e^{-L/C}\|\psi\|_L. 
\end{equation}
 As
$V^{\rm per}_\omega$ is uniformly bounded, it follows from our choice for $L$ and $l$
and estimate \eqref{br26} that, for $E$ sufficiently small,
\begin{gather*}
  \begin{split}
    \int_{\Lambda_{2l}}|\psi(x)|^2dx &\geq
    \frac1{C_1}\left(\frac{l}{L}\right)^2
    \int_{\Lambda_{2L}}|\varphi(x)|^2dx-C\|\varphi -
    \overline{e_{\theta}}\psi\|_L^2 \\ & \geq \frac1{\tilde
      C_1}\left(\frac{l}{L}\right)^2 \int_{\Lambda_{2L}}|\psi(x)|^2dx,
  \end{split}
  \\
    \int_{\Lambda_{2l}}V^{\rm per}_\omega(x)|\psi(x)|^2dx =
    \int_{\Lambda_{2l}}V^{\rm per}_\omega(x)|\varphi(x)|^2dx
    +C\|\varphi - \overline{e_{\theta}}\psi\|_L^2 \leq \tilde C_2 E
    \int_{\Lambda_{2l}}|\psi(x)|^2dx.
\end{gather*}
Hence, we obtain inequalities \eqref{411a} - \eqref{411b} with $\varphi$ replaced by $\psi \in 
P_q L^2(\R^2)$. Now, we write
$\psi = \sum_{j\geq0}a_j e_j$ (see \eqref{br1}). Using the fact that $\{e_{j}\}_{j \geq 0}$ is
an orthogonal family on any disk centered at 0 (this is due to the
rotational symmetry), we compute
\begin{equation}
  \label{eq:16}
  \int_{\Lambda_{2l}}|\psi(x)|^2dx\leq\int_{|x|\leq \sqrt{2} l}|\psi(x)|^2dx
  =\sum_{j \geq 0}|a_j|^2\int_{|x|\leq \sqrt{2} l}|e_j(x)|^2dx,
\end{equation}
and
\begin{equation}
  \label{eq:17}
  \int_{\Lambda_{2L}}|\psi(x)|^2dx \geq \int_{|x|\leq
  L}|\psi(x)|^2dx 
  =\sum_{j\geq0}|a_j|^2\int_{|x|\leq L}|e_j(x)|^2 dx.
\end{equation}
Fix $m\geq1$ and decompose $\psi =\psi_0 + \psi_m$ where 
\begin{equation}
  \label{eq:18}
 \psi_0=\sum_{j=0}^m a_je_j, \quad \quad\psi_m=\sum_{j\geq m+1}a_je_j.
\end{equation}
Our next goal is to estimate the ratio 
\begin{equation} \label{nnn}
\frac{\int_{|x|< \sqrt{2}l}|e_{j,q}(x)|^2 dx}{\int_{|x|< L}|e_{j,q}(x)|^2 dx},
\quad j \geq m + 1,
\end{equation}
where $l$, $m$, and $L$ satisfy \eqref{451} with suitable ${\cal C}$, under the hypotheses that $l$, and hence $m$ and $L$ are sufficiently large. Passing to polar coordinates $(r, \theta)$, and then changing the variable $s = \frac{b\rho^2}{2j}$ in both the numerator and the denominator of \eqref{nnn}, we find that 
\begin{equation} \label{mrs3}
\frac{\int_{|x|< \sqrt{2}l}|e_{j,q}(x)|^2 dx}{\int_{|x|< L}|e_{j,q}(x)|^2 dx} = 
\frac{\int_0^{bl^2/j} e^{-s(j-q)} s^{j-q} L_q^{(j-q)}(js)^2 ds}{\int_0^{bL^2/(2j)} 
e^{-s(j-q)} s^{j-q} L_q^{(j-q)}(js)^2 ds}. 
\end{equation}
Employing estimates \eqref{mrs1} and \eqref{mrs2}, we get 
\begin{equation} \label{mrs4}
\frac{\int_0^{bl^2/j} e^{-s(j-q)} s^{j-q} L_q^{(j-q)}(js)^2 ds}{\int_0^{bL^2/(2j)} 
e^{-s(j-q)} s^{j-q} L_q^{(j-q)}(js)^2 ds} \leq 
C(q) \left(\frac{j}{j-q}\right)^{2q} 
\frac{\int_0^{bl^2/j} e^{(j-q)f(s)}ds}{\int_0^{\epsilon(j)} 
e^{(j-q)f(s)}ds}
\end{equation}
where 
$$
f(s) : = \ln{s} - s, \quad s>0,
$$
and 
$$
\epsilon(j) = \left\{
\begin{array} {l}
\frac{1}{2} \quad {\rm if} \quad j \leq bL^2, \\
\frac{bL^2}{2j} \quad {\rm if} \quad j > bL^2.
\end{array}
\right.
$$Note that the function $f$ is increasing on the interval $(0,1)$. Since $j \geq m+1$, and 
${\cal C}$,  
the constant in \eqref{451}, is greater than one, we have $\frac{bl^2}{j} < 1$. Therefore, 
\begin{equation} \label{mrs5}
\int_0^{bl^2/j} e^{(j-q)f(s)}ds \leq \frac{bl^2}{j} e^{(j-q)f(bl^2/j)}.
\end{equation}
On the other hand, using a second-order Taylor expansion of $f$, we get 
$$
f(s) \geq f(\epsilon(j)) + \frac{s-\epsilon(j)}{\epsilon(j)} - \frac{1}{2}, \quad s \in (\epsilon(j), \epsilon(j)/2).
$$
Consequently, 
\begin{equation} \label{mrs6} 
\int_0^{\epsilon(j)} e^{(j-q)f(s)}ds \geq \int_{\epsilon(j)/2}^{\epsilon(j)} 
e^{(j-q)f(s)}ds \geq 
\frac{\epsilon(j)}{2} e^{(j-q)(f(\epsilon(j))-1))}. 
\end{equation}
Putting together \eqref{mrs4} - \eqref{mrs6}, we obtain 
\begin{equation} \label{mrs7}
  \begin{split}
    &\frac{\int_{|x|< \sqrt{2}l}|e_{j,q}(x)|^2 dx}{\int_{|x|<
        n}|e_{j,q}(x)|^2 dx} \leq 
    C(q) \frac{2bl^2}{j\epsilon(j)} \left(\frac{j}{j-q}\right)^{2q}
    \exp{((j-q)(f(bl^2/j) - f(\epsilon(j)) + 1)}
    \\&\hskip1cm \leq 
    \tilde{C}(q) \frac{2bl^2}{j\epsilon(j)}
    \left(\frac{j}{j-q}\right)^{2q} \left\{
      \begin{array} {l}
        j^q \exp{\left(-bl^2 + j\ln{(\frac{2e^{3/2}bl^2}{j})}\right)} \quad {\rm if} \quad j < bL^2, \\
        \exp{\left(-bl^2 + j\ln{(\frac{2e^2 l^2}{L^2})}\right)} \quad {\rm if} \quad j \geq bL^2.
      \end{array}
    \right.
  \end{split}
\end{equation}
Now, using the computations~\eqref{eq:16} and~\eqref{eq:17} done for
$\psi_m$, as well as \eqref{451}, we
obtain
\begin{equation}
  \label{eq:19}
  \begin{split}
  \int_{\Lambda_{2l}}|\psi_m(x)|^2dx&\leq Ce^{-bl^2/2+m\ln({\cal C}bl^2/2m)}
  \int_{\Lambda_{2L}}|\psi(x)|^2dx\\&\leq
  C_1\left(\frac{L}{l}\right)^2 e^{-bl^2/2+m\ln({\cal C} bl^2/m)}
  \int_{\Lambda_{2l}}|\psi(x)|^2dx. 
  \end{split}
\end{equation}
Plugging this into \eqref{411a} -- \eqref{411b}, and using the uniform boundedness of
$V_\omega$, we get that
\begin{gather*}
  \int_{\Lambda_{2l}}V_\omega(x)|\psi_0(x)|^2dx\leq \left(C_2
    E+C\left(\frac{L}{l}\right)^2 e^{-bl^2+m\ln({\cal C}bl^2/2m)}\right)
  \int_{\Lambda_{2l}}|\psi_ 0(x)|^2dx, \\
  2\int_{\Lambda_{2l}}|\psi_0(x)|^2dx\geq
  \left(\frac1{C_1}\left(\frac{l}{L}\right)^2 -e^{-bl^2+m\ln({\cal C}bl^2/2m)}
  \right) \int_{\Lambda_{2L}}|\psi(x)|^2dx.
\end{gather*}
Taking~\eqref{452} into consideration, this completes the proof of
Lemma~\ref{le:1}.
\end{proof}
Let us now complete the proof of Theorem \ref{t21}. Assume at first the
hypotheses of its first part. In particular, suppose that $u(x) \geq C(1 +
|x|)^{-\varkappa}$, $x \in  \rd$, with some $\varkappa > 2$, and $C>0$. Pick $\eta >
2/(\varkappa-2)$, and $\nu_0 > \max{\left\{\frac{1}{\varkappa-2}, \nu\right\}}$
where $\nu = \nu(\eta)$ is the number defined in Corollary
\ref{f31}. Finally, fix an arbitrary $\varkappa' > \varkappa$ and set 
$$
n \sim E^{-\nu_0}, \quad L = (2n+1)a/2, \quad l = E^{-\frac{1}{\varkappa'-2}},
\quad m \sim E^{-\frac{2}{\varkappa-2}}. 
$$
Then the numbers $m$, $l$, and $L$, satisfy \eqref{451} -- \eqref{452}
provided that $E>0$ is sufficiently small. Further, for any $\gamma_0 \in
l\ze^2 \cap \Lambda_{2L}$ we have 
\begin{equation}
 \label{gin7}
  \langle V^{\gamma_0}_\omega\psi,\psi\rangle_l
  \geq \sum_{\vert\gamma\vert\leq l}\omega_\gamma
  \int_{\Lambda_{2l}}u(x-\gamma)|\psi(x)|^2dx\geq
  \frac{1}{C_3} l^{-\varkappa} \sum_{\vert\gamma\vert\leq l}\omega_\gamma
  \int_{\Lambda_{2l}}|\psi(x)|^2dx
\end{equation}
with $C_3>0$ independent of $\theta$ and $E$. Hence, the probability that 
there exists $\gamma\in 2l\Z^2\cap\Lambda_{2L}$ and a non identically
vanishing function $\psi$ in the span of $\{e_j\}_{0\leq
j\leq m}$ such that~\eqref{eq:11} be satisfied, is not greater than the
probability that 
\begin{equation} \label{gin20}
  l^{-2} \sum_{\vert\gamma\vert\leq l}\omega_\gamma \leq C_3 E l^{\varkappa - 2}
  = C_3 E^{\frac{\varkappa'-\varkappa}{\varkappa'-2}}. 
\end{equation}
Applying a standard large-deviation estimate (see e.g. \cite[Subsection
  8.4]{k} or \cite[Section 3.2]{k3}), we easily find that the probability that \eqref{gin20} holds, is
bounded by 
$$
\exp{\left(C_4  l^2 \ln{\pro(\omega_0 \leq C_3
    E^{\frac{\varkappa'-\varkappa}{\varkappa'-2}})}\right)} = 
\exp{\left(C_4  E^{\frac{2}{\varkappa'-2}}\ln{\pro(\omega_0 \leq C_3
    E^{\frac{\varkappa'-\varkappa}{\varkappa'-2}})}\right)}  
$$
with $C_4$ independent of $\theta$ and $E>0$ small enough. 
Applying our hypothesis that $\pro(\omega_0 \leq E) \sim CE^{\kappa}$, $E
\downarrow 0$, with $C>0$ and $\kappa > 0$, we find that for any $\varkappa' >
\varkappa$, $\theta \in  {\mathbb T}_{2L}^*$, and sufficiently small $E>0$, we have 
\bel{gin2}
\pro(r_{q, n,\omega}(\theta)\text{ has an eigenvalue less than}\,E) \leq 
\exp{\left(-C_5 E^{\frac{2}{\varkappa'-2}} |\ln{E}|\right)}
\ee
with $C_5>0$ independent of $\theta$ and $E$. Putting together \eqref{37},
\eqref{br8} and \eqref{gin2}, and taking into account that ${\rm area}\,{\Lambda}_{2L}^* = \pi^2 L^{-2}$, we get 
$$
{\cal N}_b(2bq + E) - {\cal N}_b(2bq) \leq \frac{b}{2\pi} \exp{\left(-C_5
  E^{\frac{2}{\varkappa'-2}} |\ln{E}|\right)} + \exp{(-E^{-\eta})}
$$
which implies
$$
\liminf_{E \downarrow 0} \frac{\ln{|\ln{{\cal N}_b(2bq + E) - {\cal
	N}_b(2bq)}|}}{|\ln{E}|} \geq \frac{2}{\varkappa'-2}
$$
for any $\varkappa' > \varkappa$. Letting $\varkappa' \downarrow \varkappa$, we get
\eqref{gin22}. \\
Assume now the hypotheses of Theorem \ref{t21} ii). In particular, we suppose  
that $u(x) \geq Ce^{-C|x|^{\beta}}$, $x \in \rd$, $C>0$, $\beta > 0$. 
Put $\beta_0 = \max{\{1,2/\beta\}}$. Pick an arbitrary $\beta' > \beta$ and set 
$$
l = |\ln{E}|^{1/\beta'}, \quad m \sim |\ln{E}|^{\beta_0}.
$$
Then \eqref{451} - \eqref{452} are satisfied provided that $E>0$ is
sufficiently small, and similarly to \eqref{gin7}, 
for any $\gamma_0 \in
2l\ze^2 \cap \Lambda_{2L}$ we have 
$$
\langle V^{\gamma_0}_\omega\psi,\psi\rangle_l
  \geq
  \frac{1}{C_6} e^{-c_6 l^{\beta}}  \sum_{\vert\gamma\vert\leq l}\omega_\gamma
  \int_{\Lambda_{2l}}|\psi(x)|^2 dx
$$
with $C_6>0$ independent of $\theta$ and $E$. Arguing as in the derivation of
\eqref{gin2}, we get
\bel{gin30}
\pro(r_{q, n,\omega}(\theta)\text{ has an eigenvalue less than}\,E) \leq 
\exp{\left(-C_7 |\ln{E}|^{1+2/\beta'} \ln{|\ln{E}|}\right)}
\ee
with $C_7 > 0$ independent of $\theta$ and $E$. As in the previous case, we put together \eqref{37},
\eqref{br8} and \eqref{gin2}, and obtain the estimate 
$$
{\cal N}_b(2bq + E) - {\cal N}_b(2bq) \leq \frac{b}{2\pi} \exp{\left(-C_7
  |\ln{E}|^{1+2/\beta'} \ln{|\ln{E}|}\right)}
 + \exp{(-E^{-\eta})}
$$
which implies
$$
\liminf_{E \downarrow 0} \frac{\ln{|\ln{{\cal N}_b(2bq + E) - {\cal
	N}_b(2bq)}|}}{\ln |\ln{E}|} \geq 1 + \frac{2}{\beta'}
$$
for any $\beta' > \beta$. Letting $\beta' \downarrow \beta$, we get
\eqref{br22}. \\ 
Finally, let us assume the hypotheses of Theorem \ref{t21} iii). 
In particular, we assume that $u(x) \geq C {\bf 1}_{\{x \in \rd; |x-x_0| <
  \varepsilon\}}$ with some $C>0$, $x_0 \in \rd$, and $\varepsilon > 0$. 
Due to $\tau_{x_0} H_0 \tau_{x_0}^* = H_0$ and $\tau_{x_0} {\bf 1}_{\{x \in \rd ; |x-x_0| <
  \varepsilon\}} \tau_{x_0}^* =  {\bf 1}_{\{x \in \rd ; |x| <
  \varepsilon\}}$ we can assume without loss of generality that $x_0 = 0$. 
Our first goal is to estimate from below the ratio 
\begin{equation}
  \label{eq:23}
  R_\gamma = R_{\gamma,m,q} : = \frac{\int_{|x-\gamma|\leq\varepsilon}|{\cal
  P}_m(x)|^2 dx}
  {\int_{|x|\leq\sqrt{2}l}|{\cal P}_m(x)|^2 dx} 
\end{equation}
where 
\bel{fin30}
{\cal P}_m(x) : = \sum_{j=0}^q c_j e_{j,q}(x), \quad x \in \rd, 
\ee
with $0 \neq {\bf c} = (c_0, c_1, \ldots, c_m) \in {\mathbb C}^m$.  
\begin{lemma} \label{lfin1}
Let $q \in {\mathbb Z}_+$. Let $\pi(s) = \sum_{j=0}^q c_j s^j$, $s\in
\re$. Moreover, 
let $p \in {\mathbb Z}_+$, $\rho  \in
(0,\infty)$.
Then we have 
\begin{multline}
  \label{fin25}
    \left(\prod_{r=0}^q (r!)^2\right) \frac{e^{-(q+1)\rho
      }\rho^{q(q+1)}}{(1+\rho^q)^q}\frac{\rho^{p+1}}{(p + 2q +1)^{(q +
        1)^2 - q}}|{\bf c}|^2\\ \leq \int_0^{\rho} |\pi(s)|^2 e^{-s} s^p
    ds \leq (1 + \rho^q) \frac{\rho^{p+1}}{p+1} |{\bf c}|^2
\end{multline}
where  ${\bf c} : = (c_0,c_1,\ldots,c_q) \in {\mathbb C}^{q+1}$ and
$|{\bf c}|^2=|c_0|^2+\cdots+|c_q|^2$. 
\end{lemma}
\begin{proof}
Let ${\cal M}$ be the $(q+1) \times (q+1)$ positive-definite matrix with
entries $\int_0^{\rho} s^{j+k+p} e^{-s} ds$, $j,k = 0,1,\ldots,q$. Then we have 
$$
\int_0^{\rho} |\pi(s)|^2 e^{-s} s^p ds = \langle {\cal M} {\bf c}, {\bf c}\rangle
\leq \|{\cal M}\| |{\bf c}|^2. 
$$
Further, ${\cal M} = \int_0^\rho {\cal E}(s) ds$ where ${\cal E}(s)$, $s \in
(0,\rho)$, is the rank-one matrix with entries $s^{j+k+p} e^{-s} s^p$, $j,k =
0,1,\ldots,q$. Obviously, 
$$
\|{\cal E}(s)\| = \sqrt{\sum_{j=0}^q s^{2j}}\, e^{-s} s^p \leq (1 + s^q) e^{-s}
s^p, \quad s \in (0,\rho), 
$$
and
$$
\|{\cal M}\| \leq \int_0^\rho \|{\cal E}(s)\| ds \leq  \int_0^\rho (1 + s^q) e^{-s}
s^p ds \leq \frac{\rho^{p+1}(1+\rho^q)}{p+1}
$$
which yields the upper bound in \eqref{fin25}.  
Next, we have 
\begin{equation} \label{fin26} \frac{\det{{\cal M}}}{\|{\cal M}\|^q}
  |{\bf c}|^2 \leq \int_0^{\rho} |\pi(s)|^2 e^{-s} s^pds.
\end{equation}
Further, 
\bel{fin27}
e^{-(1+q)\rho} \det{\tilde{\cal M}} \leq \det{{\cal M}}
\ee
where $\tilde{\cal M}$ is the $(q+1) \times (q+1)$-matrix with entries
$\int_0^{\rho} s^{j+k+p} ds = \frac{\rho^{j+k+p+1}}{j+k+p+1}$, $j,k =
0,1,\ldots,q$, and 
\begin{equation} \label{fin28}
\det{\tilde{\cal M}} = \rho^{q(q+1)} \Delta_q
\end{equation}
where $\Delta_q = \Delta_q(p)$ is the determinant of the $(q+1) \times (q+1)$-matrix
with entries $(j+k+p+1)^{-1}$, $j,k = 0,1,\ldots,q$. On the other
hand, it is easy to check that
\begin{equation*}
\Delta_q =
\frac{(q!)^2}{(p + 2q + 1)\prod_{r=0}^{q-1}(p+q + r +1)^2} \Delta_{q-1}, \quad q\geq1,\
  p\geq0, \quad \Delta_0 = \frac{1}{p+1}.  
\end{equation*}
Hence, for $q\geq1$ and $p\geq0$
\begin{equation} 
\label{fin29}
\frac{\prod_{r=0}^q(r!)^2}{(p+2q+1)^{(q+1)^2}} \leq \Delta_q. 
\end{equation}
Putting together \eqref{fin26} -- \eqref{fin29} and using the upper bound in
\eqref{fin25}, we obtain the corresponding lower bound. 
\end{proof}
In the following proposition we obtain the needed lower bound of ratio
\eqref{eq:23}.
\begin{pr} \label{pfin1}
There exists a constant ${\cal C}>0$ such that for sufficiently large $m$ and $l$
ratio \eqref{eq:23} satisfies the estimates
\begin{equation}
  \label{eq:26}
  R_{\gamma} \geq e^{-{\cal C}m\ln{l}}
\end{equation}
for each linear combination ${\cal P}_m$ of the form \eqref{fin30}. 
\end{pr}
\begin{proof}
Evidently, 
\bel{fin32}
  \int_{|x-\gamma|\leq\varepsilon}|{\cal P}_m(x)|^2 dx =
\int_{|x|\leq\varepsilon}|{\cal P}_m(x+\gamma)|^2 dx = 
\int_{|x|\leq\varepsilon}|(\tau_{\gamma}{\cal P}_m)(x)|^2, 
\end{equation}
$$
  \int_{|x|\leq\sqrt{2}l}|{\cal P}_m(x)|^2 dx \leq
  \int_{|x-\gamma|\leq2\sqrt{2}l}|{\cal P}_m(x)|^2 = 
$$
\bel{fin33}
 \int_{|x|\leq2\sqrt{2}l}|{\cal P}_m(x+\gamma)|^2 dx = 
\int_{|x|\leq2\sqrt{2}l}|(\tau_{\gamma}{\cal P}_m)(x)|^2 dx,
\ee
the magnetic translation operator $\tau_{\gamma}$ being defined in
\eqref{fin34}. Using the fact that $\tau_{\gamma}$ commutes with the the
creation operator $a^*$ (see \eqref{fin10}), we easily find that \eqref{fin11}
implies 
\bel{fin12}
(\tau_{\gamma}{\cal P}_m)(x) = \sum_{j=0}^m \tilde{c}_j (a^*)^q
\left(z^j e^{\zeta z} e^{-b|z|^2/4}\right) 
\ee 
where $z = x_1 + i x_2$, $\zeta = - \frac{b}{2} (\gamma_1 -i \gamma_2)$, and the
coefficients $\tilde{c}_j$, $j=0,1,\ldots,m$, may depend on $\gamma$, $b$ and
$q$ but are independent of $x \in \rd$. Applying \eqref{br1} and
\eqref{fin11}, we get 
$$
\sum_{j=0}^m \tilde{c}_j (a^*)^q
\left(z^j e^{\zeta z} e^{-b|z|^2/4}\right) = \sum_{j=0}^m \tilde{c}_j \sum_{k=0}^{\infty}
 \frac{\zeta^k}{k!} (a^*)^q
\left(z^{j+k}  e^{-b|z|^2/4}\right) = 
$$
\bel{fin13}
e^{-b|z|^2/4} \sum_{j=0}^m \hat{c}_j z^{j-q}\sum_{k=0}^{\infty}
 \frac{(\zeta z)^k}{k!} L_q^{(j+k-q)}(b|z|^2/2) 
\ee
with $\hat{c}_j$, $j=0,1,\ldots,m$, independent of $x \in \rd$. By
\cite[Eq.(8.977.2)]{gr} we have 
\bel{fin14}
\sum_{k=0}^{\infty}
 \frac{(\zeta z)^k}{k!} L_q^{(j+k-q)}(b|z|^2/2) = 
e^{\zeta z}L_q^{(j-q)}\left(\frac{b|z|^2}{2}-\zeta z\right),
\ee
while the Taylor expansion formula entails 
\bel{fin15}
L_q^{(j-q)}\left(\frac{b|z|^2}{2}-\zeta z\right) = \sum_{s=0}^q \frac{(-\zeta
  z)^s}{s!}\frac{d^s L_q^{(j-q)}(\xi)}{d\xi^s}_{\big|\xi = b|z|^2/2}, 
\ee
and \cite[Eq.(8.971.3)]{gr} yields
\bel{fin16}
\frac{d^s L_q^{(j-q)}(\xi)}{d\xi^s} = (-1)^s L_{q-s}^{(j-q+s)}(\xi), \quad \xi
\in \re. 
\ee
Combining \eqref{fin12} - \eqref{fin16}, we find that 
\bel{fin35}
(\tau_{\gamma}{\cal P}_m)(x) = e^{\zeta z} \tilde{\cal P}_m(x), \quad x \in \rd,
\ee
where 
$$
\tilde{\cal P}_m(x) = e^{-b|z|^2/4} \sum_{j=0}^m \hat{c}_j \sum_{s=0}^q
\frac{\zeta^s}{s!} z^{j+s-q} L_{q-s}^{(j+s-q)}(b|z|^2/2) = 
$$
\bel{fin36}
e^{-b|z|^2/4} \sum_{p=0}^{m+q} z^{p-q} \phi_{p,q}(b|z|^2/2), 
\ee
and $\phi_{p,q}$, $p=0,\ldots,m+q$, are polynomials of degree not exceeding
$q$; moreover, if $p<q$, then the minimal possible degree of the non-zero
monomial terms in $\phi_{p,q}$, is $q-p$. Bearing in mind that $|e^{\zeta z}|^2
= e^{x\cdot\gamma}$ and $|\gamma| \leq \frac{\sqrt{2}}{2} l$, we find that 
there exists a constant $C$ such that for sufficiently large $l$ we have 
\bel{fin37}
R_{\gamma} \geq e^{-Cl^2} \tilde R 
\ee
where 
\begin{equation}
  \label{eq:23a}
  \tilde R=\frac{\int_{|x|\leq\varepsilon} |\tilde{\cal P}_m(x)|^2 dx}
  {\int_{|x|\leq2\sqrt{2}l}|\tilde{\cal P}_m(x)|^2dx}, 
\end{equation}
the functions $\tilde{\cal P}_m$ being defined in \eqref{fin36}. 
Passing to the polar coordinates $(r,\theta)$ in $\rd$, after that changing
the variable $s = br^2/2$, and taking into account the rotational symmetry 
we find that for each $R>0$ we have 
$$
\int_{|x| \leq R} |{\cal P}_m(x)|^2dx = \frac{2\pi}{b} \sum_{p=0}^{m+q}
\left(\frac{2}{b}\right)^{p-q} \int_0^{\rho} s^{p-q}e^{-s}|\phi_{p,q}(s)|^2 ds
= 
$$
\bel{fin20}
\sum_{p=0}^m \int_0^{\rho} s^p e^{-s} |\Pi_{p,q}(s)|^2 ds + 
\sum_{p=1}^q \int_0^{\rho} s^p e^{-s} |\tilde{\Pi}_{p,q}(s)|^2 ds; 
\ee
if $q=0$, then the second term in the last line of \eqref{fin20} should be set
equal to zero. Here $\rho = bR^2/2$, $\Pi_{p,q}(s) = \sqrt{\frac{2\pi}{b}
  \left(\frac{2}{b}\right)^p}\phi_{p+q,q}(s)$, $p = 0,\ldots,m$, $\tilde{\Pi}_{p,q} = \sqrt{\frac{2\pi}{b}
  \left(\frac{2}{b}\right)^{-p}}s^{-p}\phi_{q-p,q}(s)$, $p = 1,\ldots,q$. Note
that the degree of the polynomials $\Pi_{q,p}$ does not exceed $q$, and the
the degree of the polynomials 
$\tilde{\Pi}_{q,p}$ does not exceed $q-p$. Bearing in mind \eqref{fin20} and
applying Lemma \ref{lfin1}, we easily deduce the existence of a constant $C>0$
such that for sufficiently large $m$ and $l$ we have
$$
  \tilde R\geq e^{-Cm\ln{l}}, 
$$
which combined with \eqref{fin37} yields \eqref{eq:26}. 
\end{proof}
Next, we pick an arbitrary $\eta$ and $\nu = \nu(\eta)$, the number defined
in Corollary \ref{f31}. Further, we choose $\varsigma > 1$ and $\delta \in
(0,1/2)$ so that $\varsigma\left(1-\delta\right)>1+2\nu$, 
and set 
\begin{equation}
\label{fin31}
l = |\ln E|^{\delta/2}, \quad 
  m \sim \frac{\varsigma|\log E|}{\log|\log E|}, \quad L = (2n+1)a/2.
\end{equation}
Then, for $E$ sufficiently
small, \eqref{451} -- \eqref{452} are satisfied. Further, we impose the
additional condition that
$\mu := \frac{C\varsigma\delta}{2}<1$ 
where ${\cal C}$ is the constant
in~\eqref{eq:26}, which is compatible with the conditions on $\varsigma$ and
$\delta$ formulated above.  Now, the probability that 
there exists $\gamma\in 2l\Z^2\cap\Lambda_{2L}$ and a non identically
vanishing function $\psi$ in the span of $\{e_j\}_{0\leq
j\leq m}$ such that~\eqref{eq:11} be satisfied, is not greater than the
probability that 
\begin{equation*}
 l^{-2} \sum_{|\gamma|\leq l}\omega_\gamma\leq l^{-2} E^{1-\mu} =
 E^{1-\mu}|\ln{E}|^{\delta}. 
\end{equation*}
Arguing as in the derivation of \eqref{gin2} and \eqref{gin30}, we conclude
that for any $\theta \in {\mathbb T}_{2L}^*$ we have 
$$
    \pro(r_{q,m,\omega} \text{ has an eigenvalue less than
      }\,E)\leq 
$$
\begin{equation} \label{gin32}
\exp{\left(C_8 l^2\log\pro(\omega_0\leq E^{1-\mu}|\ln{E}|^{\delta}\right)} 
\leq \exp{\left(-C_9|\ln
      E|^{1+\delta} \ln|\ln{E}|\right)} 
\end{equation}
with positive $C_8$ and $C_9$ independent of $\theta$ and $E>0$ small
enough. Combining the upper bound in \eqref{37}, \eqref{br8}, and
\eqref{gin32}, we get \eqref{br23}. \\
This completes the proof of the upper bounds in Theorem~\ref{thin1}.

\section{Proof of Theorem \ref{thin1}: lower bounds of the IDS}
\setcounter{equation}{0}
In this section we get the lower  bounds of ${\cal N}_b(2bq+E) - {\cal
  N}_b(2bq)$ needed for the proof of Theorem
\ref{thin1}. 
\begin{theorem} \label{t22}
Assume that ${\bf H}_1$ -- ${\bf H}_4$ hold, that almost surely $\omega_{\gamma}
\geq 0$, $\gamma \in {\mathbb Z}^2$, and \eqref{fin70} is valid.  
Fix the Landau level $2bq$, $q \in {\mathbb Z}_+$.\\ 
i) 
We have 
\begin{equation} \label{gin23}
\liminf_{E \downarrow 0} \frac{\ln{|\ln{({\cal N}_b(2bq + E) - {\cal
	N}_b(2bq))}|}}{|\ln{E}|} \leq \frac{2}{\varkappa-2},  
\ee
where $\varkappa$ is the constant in \eqref{kin2}.\\
ii) Let $u(x) \leq e^{-C|x|^{\beta}}$, 
$x \in \rd$, for some $C>0$ and $\beta \in (0,2]$. Then we have 
\bel{fin65a}
\limsup_{E \downarrow 0} \frac{\ln{({\cal N}_b(2bq + E) - {\cal
      N}_b(2bq))}}{|\ln{E}|^{1+2/\beta}} \geq -  \frac{\pi \kappa}{C},  
\end{equation}
if $\beta \in (0,2)$, and
\bel{fin65}
\liminf_{E \downarrow 0} \frac{\ln{({\cal N}_b(2bq + E) - {\cal
      N}_b(2bq))}}{|\ln{E}|^2} \geq - \pi \kappa \left(\frac{2}{b} + \frac{1}{C}\right), 
\end{equation}
if $\beta = 2$. Therefore,  
\bel{hin1}
\limsup_{E \downarrow 0} \frac{\ln{|\ln{({\cal N}_b(2bq + E) -  {\cal
	N}_b(2bq))|}}}{\ln{|\ln{E}|}} \leq 1 + 2/\beta. 
\ee
\end{theorem}
Note that the combination of Theorem \ref{t21} with Theorem \ref{t22}
completes the proof of Theorem \ref{thin1}.\\
Let us prove now Theorem \ref{t22}.  Pick $\eta \geq
\frac{2}{\varkappa-2}$ in the case of its first part, or an arbitrary
$\eta > 0$ in the case of its second part.  As above, set $n \sim
E^{-\nu}$ where $\nu = \nu(\eta)$ is the number defined in Corollary
\ref{f31}, and $L = (2n+1)a/2$.  Bearing in mind the lower bound in
\eqref{37}, and \eqref{br7}, we conclude that it suffices to estimate
from below the quantity 
\begin{equation}
 \label{fin40}
  \begin{split}
{\mathbb E}\left(\rqno(E)\right) &= 
\frac{1}{(2\pi)^{2}} \int_{\Lambda_{2L}^*} {\mathbb E}(N(E;
r_{q,n,\omega}(\theta)) d\theta \\&= (2\pi)^{-2} \int_{\Lambda_{2L}^*}
\sum_{j=1}^{{\rm rank}\,r_{q,n,\omega}(\theta)} {\mathbb
  P}(\lambda_j(\theta) < E) d\theta \\&\geq
(2\pi)^{-2} \int_{\Lambda_{2L}^*}  {\mathbb P}(\lambda_1 (\theta) < E)
  d\theta.
\end{split}
\end{equation}
Fix an arbitrary $\theta  \in {\mathbb T}_{2L}^*$. Evidently, ${\mathbb
  P}(\lambda_1 (\theta) < E)$ is equal to the probability that there exists
  a non-zero function $f \in {\rm Ran}\,r_{q,n,\omega}(\theta)$ such that 
\bel{bel41}
\int_{\Lambda_{2L}} V_{\omega}(x) |f(x;\theta)|^2 dx < E \int_{\Lambda_{2L}}
  |f(x;\theta)|^2 dx.
\ee
Further, pick the trial
  function 
\bel{fin42a}
\varphi(x;\theta) = \sum_{\gamma \in
  2L{\mathbb Z}^2} e^{-i\theta(x + \gamma)}(\tau_{\gamma}\tilde{\varphi})(x),
\quad x \in \Lambda_{2L}, \quad \theta \in {\mathbb T}_{2L}^*, 
\ee
where 
\bel{fin43}
\tilde{\varphi}(x) = \tilde{\varphi}_q(x) : = \overline{z}^q e^{-b|z|^2/4}, \quad z = x_1 + i x_2, \quad 
\overline{z} = x_1 - i x_2.
\ee
Since the function $\tilde{\varphi}_q$ is proportional to $e_{0,q}$ (see
  \eqref{br1}), we have $\varphi \in {\rm
  Ran}\,r_{q,n,\omega}(\theta)$. Therefore, the probability that there exists
  a non-zero function $f \in {\rm Ran}\,r_{q,n,\omega}(\theta)$ such that 
\eqref{bel41} holds, is not less than the probability that
\bel{fin44}
\int_{\Lambda_{2L}} V_{\omega}(x) |\varphi(x;\theta)|^2 dx < E \int_{\Lambda_{2L}}
  |\varphi(x;\theta)|^2 dx.
\ee 
\begin{lemma} \label{lfin2}Let the function $\varphi$ be defined as in
  \eqref{fin42a} -- \eqref{fin43}. Then there exist $L_0 > 0$ and $c_1 > 0$ 
independent of $\theta$ such that $L \geq L_0$ implies 
\bel{fin45} 
\int_{\Lambda_{2L}}
  |\varphi(x;\theta)|^2 dx > c_1. 
\ee
\end{lemma}
\begin{proof}
We have $\varphi = \varphi_0 + \varphi_{\infty}$ where
\bel{fin49} 
\varphi_0(x;\theta) = e^{-i\theta x} \tilde{\varphi}(x), 
\ee
\bel{fin50} 
\varphi_{\infty}(x;\theta) = \sum_{\gamma \in
  2L{\mathbb Z}^2, \, \gamma \neq 0} e^{-i\theta(x +
  \gamma)}(\tau_{\gamma}\tilde{\varphi})(x). 
\ee
Note that 
\bel{fin46}
\sup_{x \in \Lambda_{2L}} |\varphi_{\infty}(x;\theta)| \leq \tilde{c}
  e^{-\tilde{c} L^2}
\ee
with $\tilde{c}$ independent of $L$ and $\theta$. Further, 
\begin{equation}
  \label{fin47}
  \begin{split}
\int_{\Lambda_{2L}}
  |\varphi(x;\theta)|^2 dx &\geq \frac{1}{2} \int_{\Lambda_{2L}}
  |\varphi_0(x;\theta)|^2 dx - 2 \int_{\Lambda_{2L}}
  |\varphi_{\infty}(x;\theta)|^2 dx \\&\geq 
\frac{1}{2} \int_{\rd}
  |\tilde{\varphi}(x)|^2 dx - 8 \tilde{c}L^2 
  e^{-\tilde{c} L^2}. 
  \end{split}
\end{equation}
Taking into account that $\int_{\rd}
  |\tilde{\varphi}|^2 dx = \frac{2\pi}{b} \left(\frac{2}{q}\right)^q q!$,
  we find that \eqref{fin47} implies \eqref{fin45}. 
\end{proof}
By assumption we have 
\bel{fin42}
u(x) \leq C v(x), \quad C>0, \quad x \in \rd,
\ee
where $v(x) : = (1+ |x|)^{-\varkappa}$ in the case of Theorem \ref{t22} i), and 
$v(x) : = e^{-C|x|^{\beta}}$ in the case of Theorem \ref{t22} ii). 
Since $\omega_{\gamma} \geq 0$, inequality \eqref{fin44} will follow from 
\bel{fin48}
\sum_{\gamma \in {\mathbb Z}^2} \omega_{\gamma} \int_{\Lambda_{2L}}
  v(x-\gamma)  |\varphi(x;\theta)|^2 dx \leq c_2 E
\ee 
where $c_2 = c_1 C^{-1}$, $C$ being the constant in \eqref{fin42}, and
$c_1$ being the constant in \eqref{fin45}. Next, we write 
$$
\sum_{\gamma \in {\mathbb Z}^2} \omega_{\gamma} \int_{\Lambda_{2L}}
  v(x-\gamma)  |\varphi(x;\theta)|^2 dx \leq 
$$
\bel{fin51}
2 \sum_{\gamma \in {\mathbb Z}^2} \omega_{\gamma} \int_{\Lambda_{2L}}
  v(x-\gamma) |\varphi_0(x;\theta)|^2 dx + 
2 \sum_{\gamma \in {\mathbb Z}^2} \omega_{\gamma} \int_{\Lambda_{2L}}
  v(x-\gamma)  |\varphi_{\infty}(x;\theta)|^2 dx 
\ee
where $\varphi_0$ and $\varphi_{\infty}$ are defined in \eqref{fin49} and
\eqref{fin50} respectively.  
\begin{lemma} \label{lgin5}
Fix $q \in \ze_+$.\\
i) Let $\varkappa > 0$, $b>0$. Then there exists a constant $c'>0$ such that for
each $y \in \rd$, $L>0$, and $\theta  \in {\mathbb T}_{2L}^*$, we have 
\bel{gin40}
\int_{\Lambda_{2L}}
(1 + |x-y|)^{-\varkappa} |\varphi_0(x;\theta)|^2 dx \leq c'(1 + |y|)^{-\varkappa}.
\ee
ii) Let $\beta \in (0,2]$, $b>0$, $C>0$. If $\beta \in (0,2)$, set $b_0 : = 
  C$. If $\beta = 2$, set $b_0 : = \frac{Cb}{2C + b}$. 
Then for each $b_1 < b_0$ there exists a constant $c''>0$ such that for
each $y \in \rd$, $L>0$, and $\theta  \in {\mathbb T}_{2L}^*$, we have 
\bel{gin41}
\int_{\Lambda_{2L}}
e^{-C|x-y|^{\beta}} |\varphi_0(x;\theta)|^2 dx \leq c''e^{-b_1|y|^{\beta}}.
\ee
\end{lemma}
We omit the proof since estimates \eqref{gin40} -- \eqref{gin41}
follow from standard simple facts concerning the asymptotics at
infinity of the convolutions of functions admitting power-like or
exponential decay, with the derivatives of Gaussian functions. In the
case of power-like decay, results of this type can be found in in
\cite[Theorem 24.1]{shu}, and in the case of an exponential decay
similar results are
contained in \cite[Lemma 3.5]{hlw1}. \\
Using Lemma \ref{lgin5}, we find that under the hypotheses of Theorem
\ref{t22} i) we have \bel{fin54} 2 \sum_{\gamma \in {\mathbb Z}^2}
\omega_{\gamma} \int_{\Lambda_{2L}} v(x-\gamma)
|\varphi_0(x;\theta)|^2 dx \leq c_3 \sum_{\gamma \in {\mathbb Z}^2}
\omega_{\gamma} (1 + \gamma|)^{-\varkappa}, \ee while under the
hypotheses of Theorem \ref{t22} ii) for each $b_1 < b_0$ we have
\bel{fin54a} 2 \sum_{\gamma \in {\mathbb Z}^2} \omega_{\gamma}
\int_{\Lambda_{2L}} v(x-\gamma) |\varphi_0(x;\theta)|^2 dx \leq c_3
\sum_{\gamma \in {\mathbb Z}^2} \omega_{\gamma} e^{-b_1|\gamma|^2},
\ee where $c_3$ is independent of $L$ and $\theta$.  Further, for both
parts of Theorem \ref{t22} we have \bel{fin55} 2 \sum_{\gamma \in
  {\mathbb Z}^2} \omega_{\gamma} \int_{\Lambda_{2L}} v(x-\gamma)
|\varphi_{\infty}(x;\theta)|^2 dx \leq c_4 L^2 e^{-\tilde{c}L^2} \ee
where $c_4$ is independent of $L$ and $\theta$, and $\tilde{c}$ is the
constant in \eqref{fin46}. Since $L \sim E^{-\nu}$, $\nu>0$, we have
\bel{fin56} c_2 E - c_4 L^2 e^{-\tilde{c}L^2} \geq \frac{c_2}{2} E \ee
for sufficiently small $E>0$. Combining \eqref{fin51} with
\eqref{fin54} -- \eqref{fin56}, and setting $c_5 = c_2/(2c_3)$, we
find that \eqref{fin48} will follow from the inequality \bel{fin61}
\sum_{\gamma \in {\mathbb Z}^2} \omega_{\gamma} c_5 (1 +
|\gamma|)^{-\varkappa} \leq c_5 E, \ee in the case of Theorem
\ref{t22} i), or from the inequality \bel{fin61a} \sum_{\gamma \in
  {\mathbb Z}^2} \omega_{\gamma} e^{-b_1|\gamma|^{\beta}} \leq c_5 E,
\quad b_1<b_0, \ee in the case of Theorem \ref{t22} ii). Now pick
$l>0$ and write \bel{fin57} \sum_{\gamma \in {\mathbb Z}^2}
\omega_{\gamma} (1 + |\gamma|)^{-\varkappa} \leq \sum_{\gamma \in
  {\mathbb Z}^2, \, |\gamma| \leq l} \omega_{\gamma} + \sum_{\gamma
  \in {\mathbb Z}^2, \, |\gamma| > l} \omega_{\gamma}
|\gamma|^{-\varkappa}, \ee \bel{fin57a} \sum_{\gamma \in {\mathbb
    Z}^2} \omega_{\gamma} e^{-b_1|\gamma|^{\beta}} \leq \sum_{\gamma
  \in {\mathbb Z}^2, \, |\gamma| \leq l} \omega_{\gamma} +
\sum_{\gamma \in {\mathbb Z}^2, \, |\gamma| > l} \omega_{\gamma}
e^{-b_1|\gamma|^{\beta}}.  \ee Evidently, for each $\varkappa' \in
(2,\varkappa)$ and $b_2 < b_1$ there exists a constant $c_6>0$ such
that \bel{fin58} \sum_{\gamma \in {\mathbb Z}^2, \, |\gamma| > l}
\omega_{\gamma} |\gamma|^{-\varkappa} \leq c_6 l^{-\varkappa'+2}, \ee
\bel{fin58a} \sum_{\gamma \in {\mathbb Z}^2, \, |\gamma| > l}
\omega_{\gamma} e^{-b_1|\gamma|^{\beta}} \leq c_6 e^{-b_2 l^{\beta}}.
\ee Fix $l$ and $c_7 \in (0,c_5)$ such that \bel{fin59} 
l^{-\varkappa'+2} = \frac{c_5 - c_7}{c_6} E \ee in the case of Theorem
\ref{t22} i), or \bel{fin59a} e^{-b_2 l^{\beta}} = \frac{c_5 -
  c_7}{c_6} E \ee in the case of Theorem \ref{t22} ii).  Putting
together \eqref{fin57} - \eqref{fin59a}, we conclude that
\eqref{fin61}, or, respectively, \eqref{fin61a} will follow from the
inequality \bel{fin62} \sum_{\gamma \in {\mathbb Z}^2, \, |\gamma|
  \leq l} \omega_{\gamma} \leq c_7 E \ee provided that $l$ satisfies
\eqref{fin59} or, respectively, \eqref{fin59a}. Set $$
N(l) : =
\#\{\gamma \in {\mathbb Z}^2 \, ; \, |\gamma| \leq l\}, $$
so that we
have \bel{fin64} N(l) = \pi l^2 (1 + o(1)), \quad l \to \infty.  \ee
Evidently, the probability that \eqref{fin62} holds, is not less than
the probability that $\omega_{\gamma} \leq c_7 E/N(l)$ for each
$\gamma \in {\mathbb Z}^2$ such that $|\gamma| \leq l$. Since the
random variables $\omega_{\gamma}$ are identically distributed and
independent, the last probability is equal to ${\mathbb P}(\omega_0
\leq c_7 E/N(l))^{N(l)}$.  Combining the above inequalities, and using
the lower bound in \eqref{37}, we get \bel{fin76} {\cal N}_b(2bq+E) -
{\cal N}_b(2bq) \geq \frac{{\rm area}\,{\Lambda}^*_{2L}}{(2\pi)^2}\,
{\mathbb P}(\omega_0 \leq c_7 E/N(l))^{N(l)} - e^{-{E^{-\eta}}}, \ee
where $l$ is chosen to satisfy \eqref{fin59} with an arbitrary
$\varkappa' \in (2,\varkappa)$ in the case of Theorem \ref{t22} i), or
to satisfy \eqref{fin59a} with an
arbitrary fixed $b_2 < b_0$ in the case of Theorem \ref{t22} ii). \\
Putting together \eqref{fin76}, \eqref{fin70}, \eqref{fin59}, and
\eqref{fin64}, we get $$
\limsup_{E \downarrow 0}
\frac{\ln{|\ln{({\cal N}_b(2bq+E) - {\cal N}_b(2bq))}|}}{|\ln E|} \leq
\frac{2}{\varkappa'-2} $$
for any $\varkappa' \in (2,\varkappa)$ such
that $\eta > \frac{2}{\varkappa'-2}$.  Letting $\varkappa' \uparrow
\varkappa$, we get
\eqref{gin23}.\\
Similarly, putting together \eqref{fin76}, \eqref{fin70},
\eqref{fin59a}, and \eqref{fin64}, we get $$
\liminf_{E \downarrow 0}
\frac{\ln{({\cal N}_b(2bq+E) - {\cal N}_b(2bq))}}{|\ln
  E|^{1+\frac{1}{\beta}}} \geq - \frac{\pi \kappa}{b_2} $$
for any
$b_2 < b_0$. Letting $$
b_2 \uparrow b_0 = \left\{
\begin{array} {l}
\frac{1}{C} \quad {\rm if} \quad \beta \in (0,2),\\
\frac{bC}{b+2C} \quad {\rm if} \quad \beta = 2,
\end{array}
\right.
$$ we get
\eqref{fin65a} -- \eqref{fin65}.\\

{\bf Acknowledgments.} The financial support of the Chilean Science
Foundation {\em Fondecyt} under Grants 1020737 and 7020737 is  
acknowledged by both authors. Georgi Raikov is sincerely grateful 
for the warm hospitality of his colleagues at the Department of Mathematics,
University of Paris 13, during his visit in 2004, when a
considerable part of this work was done. 
He would  like to thank also Werner Kirsch, Hajo
Leschke and Simone Warzel for several illuminating discussions.\\

{\sc Fr{\'e}d{\'e}ric Klopp}\\
D{\'e}partement de  math{\'e}matiques\\
Universit{\'e}  de  Paris Nord\\
Avenue  J. Baptiste  Cl{\'e}ment\\
93430 Villetaneuse, France\\
E-mail: klopp@math.univ-paris13.fr\\

{\sc Georgi Raikov}\\
Departamento de Matem{\'a}ticas\\
Facultad de Ciencias\\
Universidad de Chile\\
Las Palmeras 3425\\
Santiago, Chile\\
E-mail: graykov@uchile.cl
\end{document}